\documentclass[prd,twocolumn,reprint,preprintnumbers,nofootinbib,superscriptaddress,longbibliography]{revtex4-2}

\usepackage{float}
\usepackage{bm,amsmath,amssymb}
\usepackage{booktabs}
\usepackage{graphicx}
\usepackage{xcolor}
\usepackage[colorlinks=true,allcolors=blue]{hyperref}
\usepackage{braket}
\usepackage{placeins}
\usepackage{comment}
\usepackage{multirow}

\newcommand{\GN}{Gross-Neveu}
\newcommand{\Nf}{N_f}
\newcommand{\LD}{L_D}
\newcommand{\hc}{\mathrm{H.c.}}

\begin{document}

\author{Vivan Bhatt}
\email{vivanbhatt@ku.edu}
\affiliation{Department of Physics and Astronomy, University of Kansas, Lawrence, KS 66045, USA}

\author{Zhongtian Dong}
\email{cdong@pku.edu.cn}
\affiliation{Center for High Energy Physics, Peking University, Beijing 100871, China}

\author{Kyoungchul Kong}
\email{kckong@ku.edu}
\affiliation{Department of Physics and Astronomy, University of Kansas, Lawrence, KS 66045, USA}

\author{Kwangmin Yu}
\email{kyu@bnl.gov}
\affiliation{Computational Science Department, Brookhaven National Laboratory, Upton, New York 11973, USA}

\title{Symmetry-Preserving and Scalable Variational Circuits for the Lattice Gross-Neveu Model}
\date{\today}

\begin{abstract}
We study variational preparation of ground states and sector-resolved excited states in the $(1+1)$-dimensional lattice Gross-Neveu model. We construct fixed circuits preserving either total fermion number or every flavor number, and a problem-specific scalable-circuit adaptive variational quantum eigensolver (SC-ADAPT-VQE). Classical circuit benchmarks cover massless and massive systems through 30 qubits, using exact-diagonalization and tensor-network reference results. Imposing flavor-number conservation improves the achieved energies in the completed two-flavor comparisons. Selected massive fixed-ansatz benchmarks with one to four flavors have relative energy errors of approximately $1 \times 10^{-5}$ - $2\times10^{-4}$, smaller than the quoted literature errors where available, although discretizations and computational resources differ. The adaptive pool combines translated bulk and boundary-sensitive generators with an interaction-sensitive cross-flavor family. Selected massive SC-ADAPT benchmarks with one to three flavors have relative energy errors below $2.1\times10^{-3}$. For one flavor, matrix-product-state evaluations extend the transferred circuits to 60-400 qubits without target-side operator selection or parameter optimization, giving energy deviations of approximately $0.35\%$-$1.76\%$ from density-matrix renormalization-group reference values at maximum bond dimension 256. At fixed flavor number, range caps, and append count, the adaptive construction has a precompilation local-factor count linear in volume and a volume-independent shared-parameter count; fixed-accuracy performance at arbitrary volumes remains unestablished. Sequential deflation also prepares low-lying states within a fixed flavor-number sector, with projector fidelities providing a basis-independent assessment of degenerate target eigenspaces.
\end{abstract}

\maketitle

\section{Introduction}
\label{sec:introduction}

The $(1+1)$-dimensional \GN{} model describes $\Nf$ relativistic fermion flavors coupled by a local four-fermion interaction. In the infinite-volume massless theory, asymptotic freedom and dimensional transmutation accompany dynamical mass generation associated with spontaneous breaking of a discrete chiral symmetry \cite{gross1974dynamical,thies2006grossNeveuPhaseDiagram}. Although it has no dynamical gauge field and does not reproduce the full structure of quantum chromodynamics, the model provides a controlled setting for nonperturbative fermion dynamics. Matrix-product-state studies have characterized its chiral symmetry breaking, mass-gap scaling, entanglement structure, and relativistic continuum behavior \cite{haegeman2010variationalQFT,roose2021latticeGN}. These features motivate its use as a benchmark for interacting-fermion state preparation. Here we study preparation accuracy for finite-lattice Hamiltonians at fixed lattice spacing.

Reliable preparation of interacting low-energy states is a prerequisite for Hamiltonian calculations of correlation functions, real-time evolution, and scattering. Direct Monte Carlo treatments of real-time path integrals encounter oscillatory phases \cite{alexandru2016realtime}, and Euclidean calculations at finite fermion density can suffer from sign problems \cite{troyer2005computational}. Tensor-network methods remain highly effective in one dimension \cite{takeda2015grassmannGN}, but quantum simulation represents unitary time evolution directly and therefore motivates circuit constructions for the vacuum and low-lying spectrum \cite{bauer2023fundamental}. Finite systems accessible to classical reference calculations provide benchmarks for assessing these constructions.

Several Gross-Neveu state-preparation proposals have treated the interacting vacuum as an input to a subsequent dynamical calculation. Adiabatic vacuum preparation was incorporated into a fermionic quantum-field-theory scattering algorithm \cite{jordan2014fermionic}; later proposals sought to reduce this cost by compiling classically obtained matrix-product states or by preparing gapped fermionic ground states site by site \cite{moosavian2018faster,moosavian2019sitebysite}. Other quantum studies have focused on real-time correlators, propagators, scattering observables, and hardware-aware multi-flavor dynamics \cite{li2022partonicGN,li2024scatteringGN,chowdhury2026utilityScaleGN}. Together, these developments motivate variational circuits that incorporate the exact charges and interaction structure of low-energy states from the outset.

The closest circuit-level ground-state studies use different finite-lattice formulations and circuit architectures. A four-qubit hardware-efficient Variational Quantum Eigensolver (VQE) \cite{tilly2022variational} calculation treated the massless two-flavor model and compared energies and computational-basis probabilities with exact-diagonalization and density-matrix renormalization-group results \cite{asaduzzaman2022quantumGN}. Its displayed circuit does not preserve total or flavor-resolved fermion number by construction, and its finite-lattice interaction differs from the Dirac-cell scalar-density-square interaction used here. More recently, adaptive variational quantum imaginary-time evolution was applied to massive Gross-Neveu and Thirring models for $\Nf=1,2,3,4$ and registers through 20 qubits \cite{bakalov2026massiveGN}. The present fixed-ansatz calculations achieve smaller relative energy errors than the quoted benchmarks at the available comparison points, with each error measured against the reference energy of its respective finite-lattice Hamiltonian. Ref.~\cite{bakalov2026massiveGN} uses a na\"ive two-component Dirac discretization, whereas we use staggered fermions; the discretizations and computational resources differ, so these numbers compare reported benchmark errors rather than establish a matched algorithmic ranking.

The lattice Hamiltonian considered here commutes with every flavor number $N_b$ and consequently with the total number $N=\sum_bN_b$. With a reference state of definite charge, symmetry-preserving circuits restrict the variational search to the corresponding conserved-charge sector \cite{gard2020symmetry}. A total-number-conserving ansatz permits occupation transfer among flavors, whereas a flavor-number-conserving ansatz forbids such transfer while retaining diagonal cross-flavor correlations. Preserving every $N_b$ enforces mutually commuting Abelian charges but does not by itself impose the full non-Abelian flavor symmetry. Our two fixed families share the circuit graph, layer organization, and local-phase parametrization. Removing the cross-flavor exchange parameters imposes individual flavor-number conservation. Under the optimization protocol used here, this restriction yields smaller final energy errors in the completed two-flavor comparisons, demonstrating a practical benefit of the stronger charge constraint for these benchmarks.

Adaptive construction addresses the complementary problem of selecting the circuit architecture. ADAPT-VQE grows the ansatz by ranking candidate generators through energy gradients and reoptimizing after each selection \cite{grimsley2019adapt}. SC-ADAPT-VQE organizes operators into translated bulk sums and boundary-sensitive surface corrections; for the Schwinger model, structures determined on classically tractable lattices were used to construct circuits for larger systems \cite{PRXQuantum.5.020315}. We adapt this approach to the Gross-Neveu Hamiltonian using a compact, range-truncated pool. In addition to mass-hopping bulk and surface families, it contains an interaction-sensitive family obtained from the cross-flavor component of the interaction-hopping commutator. Every retained generator preserves each $N_b$ and has a local flavor-number-preserving circuit realization. At fixed flavor number, range caps, and append count, the precompilation local-factor count grows linearly with volume while the shared-parameter count remains fixed.

We benchmark the circuits using classical simulations in massless and massive regimes, with reference energies and states obtained using the density-matrix renormalization group (DMRG), a tensor-network method, and exact diagonalization for the small-system spectral comparisons. Selected massive flavor-number-conserving fixed-ansatz benchmarks span $\Nf=1,2,3,4$ and registers through 30 qubits, with relative energy errors of approximately $1\times10^{-5}$-$2\times10^{-4}$. Selected massive SC-ADAPT benchmarks with $\Nf=1,2,3$ have relative energy errors below $2.1\times10^{-3}$ through 30 qubits. The complete one-flavor sequence is unchanged over $\LD=12$-15. In a retrospective test, freezing its ten-generator $\LD=12$ sequence and tabulated four-decimal parameters gives relative energy errors of $1.97\times10^{-3}$-$2.03\times10^{-3}$ at $\LD=13$-15 without target-side selection or reoptimization. This establishes finite-volume energy transfer for the specified rounded circuit. Matrix-product-state (MPS) evaluations extend the one-flavor study to $\LD=30,50,100,200$, corresponding to 60-400 qubits, using frozen source parameters for the first three volumes and a mixed extrapolation and frozen-parameter prescription for the last. At maximum bond dimension 256, their energies differ from the DMRG references by approximately $0.35\%$-$1.76\%$. These finite-bond evaluations retain numerical error, and fixed-accuracy performance at arbitrary volumes remains to be tested.

Following a charge-preserving sequential-deflation strategy \cite{Chandani:2024xza}, we also prepare the lowest four states by within-sector rank $k=0,1,2,3$ in the $(N_0,N_1)=(2,2)$ sector of massless and massive eight-qubit systems. The massive $k=3$ target belongs to a degenerate global eigenspace rather than being the global third excited state. A projector onto the complete eigenspace supplies a basis-independent fidelity diagnostic. These benchmarks distinguish state preparation within a conserved-charge sector from ordering in the complete spectrum.

Section~\ref{sec:model} defines the Hamiltonian and qubit conventions. Sections~\ref{sec:fixed_vqe} and~\ref{sec:sc_adapt_pool} present the fixed and adaptive circuits, respectively. Ground-state benchmarks and the transfer tests appear in Sec.~\ref{sec:numerical_results}, followed by sector-resolved excited states in Sec.~\ref{sec:excited_states} and conclusions in Sec.~\ref{sec:conclusions}.

\section{Continuum and Lattice Hamiltonians, Symmetries, and Qubit Conventions}
\label{sec:model}

\subsection{Continuum Normalization and Staggered Hamiltonian}

The massless \GN{} model is defined in Ref.~\cite{gross1974dynamical}. We study its bare-mass deformation for $\Nf$ interacting Dirac-fermion flavors in $1+1$ dimensions, with continuum normalization
\begin{equation}
  \begin{split}
    \mathcal{L}
    ={}&
    \sum_{b=0}^{\Nf-1}
    i\bar\psi_b\gamma^\mu\partial_\mu\psi_b
    -m\sum_{b=0}^{\Nf-1}\bar\psi_b\psi_b
    \\
    &+\frac{g^2}{2}
    \left(
      \sum_{b=0}^{\Nf-1}\bar\psi_b\psi_b
    \right)^2 .
  \end{split}
  \label{eq:continuum_lagrangian}
\end{equation}
Here $m$ and $g^2$ are the bare mass and four-fermion coupling, respectively. At $m=0$, the action is invariant under the discrete chiral transformation $\psi_b\to\gamma^5\psi_b$, $\bar\psi_b\to-\bar\psi_b\gamma^5$, with $\gamma^5=\gamma^0\gamma^1$; the bare mass term explicitly breaks this symmetry.

We discretize space with staggered fermions while retaining continuous time~\cite{kogut1975hamiltonian}. The open chain contains $2\LD$ staggered sites, $j=0,\ldots,2\LD-1$, separated by the lattice spacing $a$. Sites $2n$ and $2n+1$ represent the two components associated with Dirac cell $n$. The explicit staggered spinor map is given in Eq.~\eqref{eq:staggered_mapping} of Appendix~\ref{app:staggered_jw}.

We retain two phase-equivalent hopping conventions. In convention $I$, the staggered fields $\chi_{j,b}$ give the current-form hopping term
\begin{equation}
  \begin{split}
  H_{\mathrm{hop}}^{(I)}
  ={}&-\frac{i}{2a}
  \sum_{j=0}^{2\LD-2}
  \sum_{b=0}^{\Nf-1}
  \left(
    \chi_{j,b}^{\dagger}\chi_{j+1,b}
    -\chi_{j+1,b}^{\dagger}\chi_{j,b}
  \right).
  \end{split}
\end{equation}
The local number rephasing $\chi_{j,b}=i^j c_{j,b}$ converts it to the uniform real-hopping convention $R$,
\begin{equation}
  \begin{split}
  H_{\mathrm{hop}}^{(R)}
  ={}&\frac{1}{2a}
  \sum_{j=0}^{2\LD-2}
  \sum_{b=0}^{\Nf-1}
  \left(
    c_{j,b}^{\dagger}c_{j+1,b}
    +c_{j+1,b}^{\dagger}c_{j,b}
  \right).
  \end{split}
  \label{eq:hopping_hamiltonian}
\end{equation}
The rephasing is diagonal in the occupation basis and therefore leaves all number operators, the mass term, and the interaction unchanged. The two representations may consequently be written as
\begin{equation}
  H^{(\alpha)}
  =H_m+H_{\mathrm{hop}}^{(\alpha)}+H_{\mathrm{int}},
  \qquad
  \alpha\in\{I,R\}.
  \label{eq:hamiltonian_decomposition}
\end{equation}
With $\hat n_{j,b}=\chi_{j,b}^{\dagger}\chi_{j,b}=c_{j,b}^{\dagger}c_{j,b}$, the common staggered mass term is
\begin{equation}
  H_m
  =\frac{m}{2}
  \sum_{j=0}^{2\LD-1}
  \sum_{b=0}^{\Nf-1}
  (-1)^j\left(2\hat n_{j,b}-1\right).
  \label{eq:mass_hamiltonian}
\end{equation}
The four-fermion interaction is
\begin{equation}
  \begin{aligned}
    H_{\mathrm{int}}
    &=-\frac{g^2}{2a}
    \sum_{n=0}^{\LD-1}D_n^2,
    \\
    D_n
    &=\sum_{b=0}^{\Nf-1}
    \left(
      \hat n_{2n,b}-\hat n_{2n+1,b}
    \right).
  \end{aligned}
  \label{eq:interaction_hamiltonian}
\end{equation}
The operator $D_n$ is the cellwise staggered scalar-density difference. Its expansion contains both same-flavor terms and cross-flavor density-density couplings, as displayed in Eq.~\eqref{eq:dn_expansion} of Appendix~\ref{app:staggered_jw}. These terms couple flavor densities without transferring occupation between flavors. Accordingly,
\begin{equation}
  \begin{aligned}
    N_b&=\sum_{j=0}^{2\LD-1}\hat n_{j,b},
    &\qquad [H,N_b]&=0,
    \\
    N&=\sum_{b=0}^{\Nf-1}N_b .
  \end{aligned}
\end{equation}
The exact conservation of every $N_b$, and therefore of the total number $N$, underlies the two symmetry-preserving fixed variational families and the SC-ADAPT pool constructed below.

\subsection{Qubit Layout and Jordan-Wigner Convention}
\label{sec:jw_convention}

One qubit is assigned to each staggered fermionic mode. The register size and site-major register index are
\begin{align}
  N_q&=2\LD\Nf,
  \label{eq:nqubits}\\
  q(j,b)&=j\Nf+b.
  \label{eq:qubit_ordering}
\end{align}
The function $q(j,b)$ specifies the site-major register index; it does not define the algebraic Jordan-Wigner parity path. Following the general Jordan-Wigner transformation~\cite{jordan1928paulische}, we choose this path to advance along the staggered-site coordinate $j$ at fixed flavor $b$, with a Klein factor enforcing anticommutation between different flavor chains. Thus the parity convention is flavor-chain ordered even though the qubit labels are site-major. In particular, it is not the flattened Jordan-Wigner ordering $q=0,1,\ldots,N_q-1$. The explicit convention-specific transformation is given in Eq.~\eqref{eq:generic_jw_string} of Appendix~\ref{app:staggered_jw}.

For $\alpha\in\{I,R\}$, let $P_q^{(\alpha)}$, with $P\in\{X,Y,Z\}$, denote the Pauli operators in convention $\alpha$. The on-site $Z$ rotations implementing $\chi_{j,b}=i^jc_{j,b}$ leave $Z_q$ unchanged and map the nearest-neighbor hopping strings as
\begin{equation}
  \begin{aligned}
    H_{\mathrm{hop}}^{(I)}
    &\longleftrightarrow
    \frac{XY-YX}{4a},
    \\
    H_{\mathrm{hop}}^{(R)}
    &\longleftrightarrow
    \frac{XX+YY}{4a},
  \end{aligned}
  \label{eq:hopping_pauli_comparison}
\end{equation}
with the site and flavor sums understood. Their complete Pauli forms, together with the mass term, are given in Appendix~\ref{app:staggered_jw}. Convention superscripts on $X$ and $Y$ are suppressed below when the choice is unambiguous.

The interaction is obtained from
\begin{equation}
  \begin{aligned}
    D_n
    &=\frac{1}{2}
    \sum_{b=0}^{\Nf-1}
    \left[
      Z_{q(2n+1,b)}-Z_{q(2n,b)}
    \right],
    \\
    E_{\mathrm{const}}^{(\mathrm{int})}
    &=-\frac{g^2\LD\Nf}{4a}.
  \end{aligned}
\end{equation}
The second line is the identity contribution generated by the same-flavor terms in $D_n^2$. It must be retained or subtracted consistently when absolute variational energies are compared with exact diagonalization.

Throughout the fixed-ansatz calculations, we use the current-hopping convention $I$. The SC-ADAPT construction and its numerical results use the uniform real-hopping convention $R$. These conventions are related by the local rephasing $\chi_{j,b}=i^j c_{j,b}$; they use the same site-major register labels and the same flavor-chain parity convention.

\section{Symmetry-Preserving Fixed Variational Ans\"atze}

\label{sec:fixed_vqe}

Traditional VQEs in the literature often utilize universal ans\"atze such as the hardware-efficient ansatz \cite{kandala2017hardware} used in Ref.~\cite{asaduzzaman2022quantumGN}. While these ans\"atze have generally high expressivity, they do not respect the symmetries of the Hamiltonian and require little physics input. The hardware-efficient circuit considered here uses $SU(2)$ blocks connecting qubits along the Hamiltonian graph. However, the lattice Hamiltonian conserves the flavor charges \(N_b\) and their sum \(N\), as defined in Sec.~\ref{sec:model}, which are both violated by the efficient $SU(2)$ construction. We thus construct two fixed variational families that enforce these symmetries at different levels. The total-number-conserving family fixes \(N\) but permits occupation transfer among flavors, whereas the flavor-number-conserving family fixes every \(N_b\) separately.

\subsection{Reference sector and local blocks}

\label{sec:fixed_local_blocks}

Both families begin from the odd-site-filled computational-basis state
\begin{equation}
  \begin{aligned}
    |\Phi_0\rangle
    ={}&
    \bigotimes_{j=0}^{2\LD-1}
    \bigotimes_{b=0}^{\Nf-1}
    \bigl|n_{j,b}^{(0)}\bigr\rangle_{q(j,b)},
    \\
    n_{j,b}^{(0)}
    ={}&
    \frac{1-(-1)^j}{2}.
  \end{aligned}
  \label{eq:fixed_reference}
\end{equation}
It satisfies \(N_b|\Phi_0\rangle=\LD|\Phi_0\rangle\) for every flavor and \(N|\Phi_0\rangle=\Nf\LD|\Phi_0\rangle\). This occupation-basis definition is invariant under the local rephasing that relates hopping conventions \(I\) and \(R\). The fixed constructions and benchmarks use convention \(I\); accordingly, the convention-suppressed \(X\) and \(Y\) operators below denote the convention-\(I\) Pauli axes introduced in Sec.~\ref{sec:jw_convention}.

For a Pauli string \(P\), we use
\begin{equation}
  R_P(\vartheta)\equiv e^{-i\vartheta P},
  \label{eq:fixed_pauli_rotation}
\end{equation}
and write products in operator order, so that the rightmost factor acts first. Let \(u\) and \(v\) denote arbitrary qubit indices.

The elementary total-number-conserving block is
\begin{equation}
  \begin{aligned}
    B_{uv}(\alpha,\beta,\gamma)
    ={}&
    R_{Z_v}(-\gamma)R_{Z_u}(\gamma)
    \\
    &\times
    R_{Z_uZ_v}(\beta)
    \\
    &\times
    R_{Y_uY_v}(\alpha)R_{X_uX_v}(\alpha).
  \end{aligned}
  \label{eq:flavor_number_block}
\end{equation}
The tied \(XX\) and \(YY\) rotations generate number-preserving exchange, while the \(ZZ\) and local \(Z\) rotations are diagonal in the occupation basis. On a same-flavor pair, the exchange redistributes occupation within one flavor. On a cross-flavor pair, it can exchange \(|10\rangle\) and \(|01\rangle\), preserving the total number while changing the two flavor occupations separately. The detailed conservation algebra is given in Appendix~\ref{app:fixed_ansatze}.

To preserve every \(N_b\), off-diagonal exchange is restricted to pairs from the same flavor. For such a pair, cross-flavor correlations are introduced without occupation transfer through
\begin{equation}
  \begin{aligned}
    D_{uv}(\gamma,\delta)
    &=
    R_{Z_v}(-\gamma)R_{Z_u}(\gamma)
    R_{Z_uZ_v}(\delta),
    \\
    u&=q(j,b),\qquad
    v=q(k,c),\qquad b\ne c .
  \end{aligned}
  \label{eq:flavor_diagonal_block}
\end{equation}
The complete \(B\) and \(D\) blocks commute with every flavor charge for the qubits that they act upon. The local phases are anti-tied and generate the relative direction \(Z_u-Z_v\). This tying is an architectural choice rather than a consequence of flavor or number conservation, because \(Z_u\) and \(Z_v\) separately commute with all occupation-number operators.

For arbitrary variational parameters, the total-number circuit remains in the sector \(N=\Nf\LD\) but may leave the individual sectors \(N_b=\LD\). The flavor-number circuit remains in the joint sector \(N_b=\LD\) for every \(b\). These restrictions are enforced by the circuit blocks themselves and require neither penalty terms nor postselection.

\subsection{Benchmark circuit architectures}

\label{sec:fixed_architecture}

\begin{figure}[t]
    \centering
    \includegraphics[width=\linewidth]{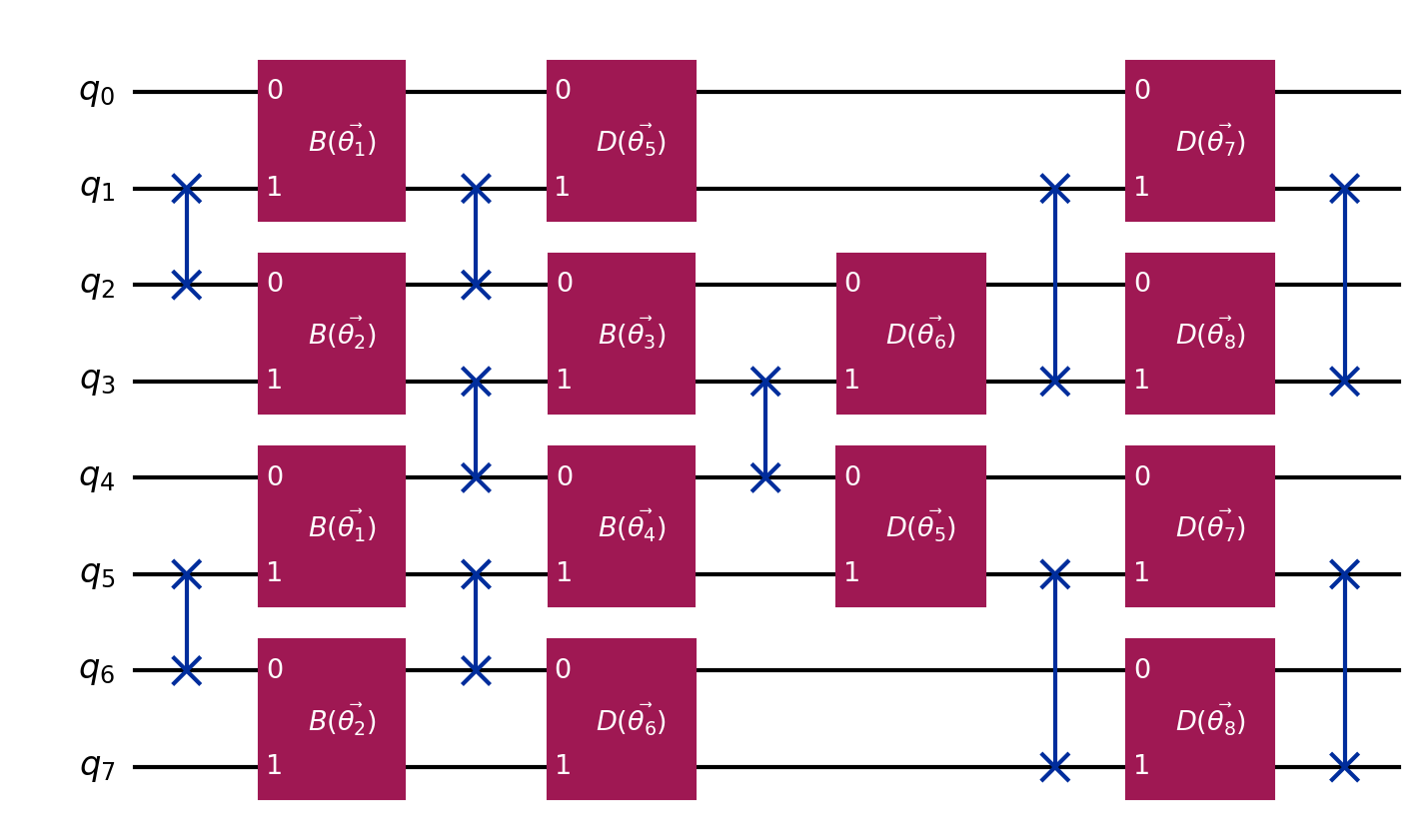}
    \caption{Example circuit for the flavor-number-conserving ansatz, with $\Nf=2$ and $\LD=2$. For the total-number-conserving ansatz, the $D$ blocks that act on off-diagonal qubit pairs are also replaced by the $B$ blocks that redistribute occupation across different flavors while preserving the total number.}
    \label{fig:circuit_graph}
\end{figure}

One layer contains translated blocks on every same-flavor nearest-neighbor bond, a second same-flavor block on every intracell pair, and four intracell cross-flavor topologies for each unordered flavor pair, as seen in Fig. \ref{fig:circuit_graph}. A parameter tuple is shared over translations of a fixed flavor, edge class, and cross-flavor topology, while distinct layers receive independent tuples. The total-number family uses \(B\) on all three edge classes; the flavor-number family uses \(B\) on the same-flavor classes and \(D\) on the cross-flavor classes. The total number of blocks for a single layer of this graph is given by $N_{\text{blocks}} = 2L_DN_f^2 + L_DN_f - N_f$. Although the total number of gates is highly dependent on the implementation of the entangling Pauli exponentials within the native architecture of the hardware, the scheduled Pauli rotation count can be used to provide an estimate of the resource count. Specifically, for the flavor-number-conserving ansatz, we have $N_{\text{Pauli strings}} = 6L_DN_f^2 + 9L_DN_f - 5N_f$ per layer, which linearly depends on the number of lattice sites and quadratically on the number of flavors. The quadratic flavor number dependence is also observed in the parameter count $N_{\text{parameters}} = 4N_f^2 + 2N_f$ per layer, which remains independent of the number of lattice sites. Following a brickwall structure, all even-start bonds are applied first and all odd-start bonds second; the two parity sweeps share the same tuple for a given flavor. The additional intracell stage follows these nearest-neighbor stages. Full edge-set, schedules, and ordered-product definitions are collected in Appendix~\ref{app:fixed_ansatze}.

The two constructions thus match in their circuit graph, local-phase parametrization, nearest-neighbor product order, circuit depth, and block count. The parameter and Pauli string counts differ slightly due to the difference in their cross-flavor gate content, which is the source of the stronger flavor-number constraint. Restricting optimization to the target flavor-number sector is consistent with the improved energies achieved by the flavor-number-conserving ansatz.

Finally, ``flavor-number-conserving'' refers here to exact preservation of the mutually commuting Abelian charges \(\{N_b\}\). It does not imply enforcement of the full non-Abelian flavor symmetry, commutation with generators that mix flavor labels, or restriction to a specified irreducible representation of the full flavor group.

\section{SC-ADAPT Construction}

\label{sec:sc_adapt_pool}

The SC-ADAPT ansatz follows the append-and-reoptimize procedure of ADAPT-VQE \cite{grimsley2019adapt} and the volume-surface organization of Ref.~\cite{PRXQuantum.5.020315}, extended here to the multi-flavor \GN{} Hamiltonian. Open boundaries break exact lattice-translation invariance, while translated volume sums capture repeated bulk structure and separate surface families supply boundary-local directions. The ansatz uses the reference state in Eq.~\eqref{eq:fixed_reference}
and the uniform real-hopping convention (R).

For an implemented factorized candidate $\widetilde U_i(\theta)$, the raw selection metric is the magnitude of the centered append-point energy score
\begin{equation}
  \begin{aligned}
    g_i^{\mathrm{FD}}
    &=
    \frac{E_i(\delta)-E_i(-\delta)}{2\delta},
    \\
    \delta&=10^{-3},
    \qquad
    E_i(\theta)&=
    \langle\psi|\widetilde U_i^{\dagger}(\theta)
    H\widetilde U_i(\theta)|\psi\rangle .
  \end{aligned}
  \label{eq:adapt_finite_difference}
\end{equation}
At each append step, we select the candidate with the largest $|g_i^{\mathrm{FD}}|$ and then reoptimize all circuit parameters. Selected operators remain available for subsequent reuse. Because $g_i^{\mathrm{FD}}$ estimates the derivative with respect to the implemented circuit coordinate, its magnitude depends on candidate normalization, including the factor $\eta_{gh}$ in the interaction-hopping family. The pool caps, candidate normalizations, and fixed factor orders are consequently part of the algorithmic specification rather than conventions under which the raw ranking is invariant.

\subsection{Gross-Neveu Volume-Surface Pool}

\label{sec:scalable_blocks}

For compactness, introduce the Hermitian same-flavor current
\begin{equation}
  J_{j,b}^{(d)}
  =
  i\left(
    c_{j,b}^{\dagger}c_{j+d,b}
    -c_{j+d,b}^{\dagger}c_{j,b}
  \right).
  \label{eq:sc_local_current}
\end{equation}
The four pool families are the mass-hopping volume generator
\begin{equation}
  \begin{aligned}
    O_{mh}^V(d)
    &\equiv i[\Theta_m^V,\Theta_h^V(d)]
    \\
    &=
    \sum_{j=0}^{2\LD-1-d}
    \sum_{b=0}^{\Nf-1}
    (-1)^jJ_{j,b}^{(d)},
    \qquad d\ \text{odd},
  \end{aligned}
  \label{eq:omh_definition}
\end{equation}
which vanishes for even \(d\), two boundary-local counterparts \(O_{mh}^{S0}(d)\) and \(O_{mh}^{S1}(d)\), and the \GN{}-specific interaction-hopping volume family
\begin{equation}
  O_{gh}^V(d)
  \equiv
  \left.
  i[\Theta_g^V,\Theta_h^V(d)]
  \right|_{\mathrm{cross}}.
  \label{eq:ogh_definition}
\end{equation}
Here the parent operators \(\Theta_m^V\), \(\Theta_h^V(d)\), and \(\Theta_g^V\), the surface endpoint formulas, and all normalizations are defined in Appendix~\ref{app:sc_adapt_details}. The cross-flavor projection retains terms in which a current of flavor \(\beta\) is multiplied by a density difference \(\Delta_{n,b}=\hat n_{2n,b}-\hat n_{2n+1,b}\) of a distinct flavor \(b\ne\beta\). Every retained generator commutes with every \(N_b\). This is an additional property of the selected \GN{} pool, not the defining bulk-surface principle of SC-ADAPT. Unlike \(O_{mh}^V(d)\), the interaction-hopping commutator need not vanish for even \(d\); restricting it to odd \(d\) is a pool-design choice.

\subsection{Geometric Family and Numerical Pool}

\label{sec:minimal_pool}

The default numerical pool is
\begin{equation}
  \begin{aligned}
    O_{mh}^{V}:&\quad d=1,3,5,
    \\
    O_{mh}^{S0}:&\quad d=1,3,
    \\
    O_{mh}^{S1}:&\quad d=1,3,
    \\
    O_{gh}^{V}:&\quad d=1,3,
  \end{aligned}
  \label{eq:numerical_pool}
\end{equation}
with unavailable distances omitted on short chains. The full geometric ranges are given in Appendix~\ref{app:sc_adapt_details}. These caps and the omission of interaction-hopping surface operators are numerical truncation choices, not completeness or minimality statements.

The candidate circuits use the \(R_\pm\) and X-pattern circuit convention of Fig.~3 in Ref.~\cite{PRXQuantum.5.020315}. Each local current block and each controlled-\(Z\)-dressed current block is exact, whereas exponentials of the volume or surface sums are represented by fixed first-order ordered products. The local identities, family-specific normalizations, and complete schedules are specified in Appendix~\ref{app:sc_adapt_details}.

For fixed $\Nf$ and fixed distance caps, the number of translated local factors in each volume candidate grows linearly with $\LD$, while the surface families remain boundary local. At fixed append count, the full circuit therefore has linear precompilation local-factor growth and a volume-independent number of shared variational coordinates. This defines the scalability of the circuit construction used here. It does not determine the append count or compiled depth required at fixed accuracy, or the end-to-end cost of classical simulation and optimization.

\section{Ground-State Numerical Results}
\label{sec:numerical_results}

\subsection{Numerical Setup}
\label{sec:numerical_setup}

\begin{figure}[t]
    \centering
    \includegraphics[width=\linewidth]{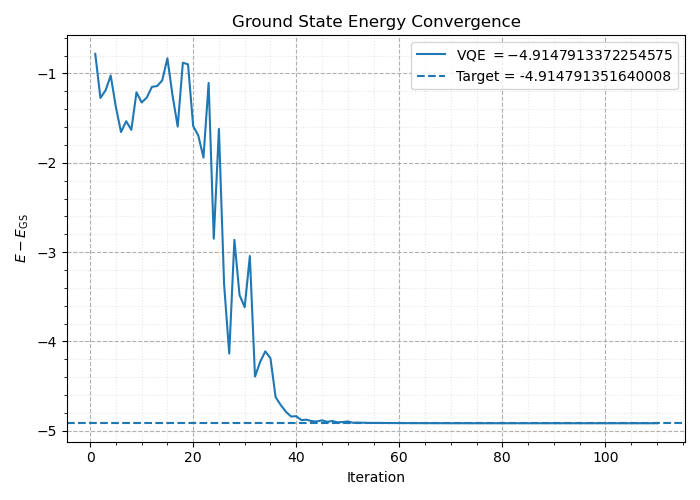}
    \caption{Ground-state energy convergence for 8-qubit flavor-number-conserving ansatz with $m=0$, $g^2=1.0$, $\Nf=2$, $\LD=2$, and open boundary conditions.}
    \label{fig:gs_energy}
\end{figure}

\begin{figure}[t]
    \centering
    \includegraphics[width=\linewidth]{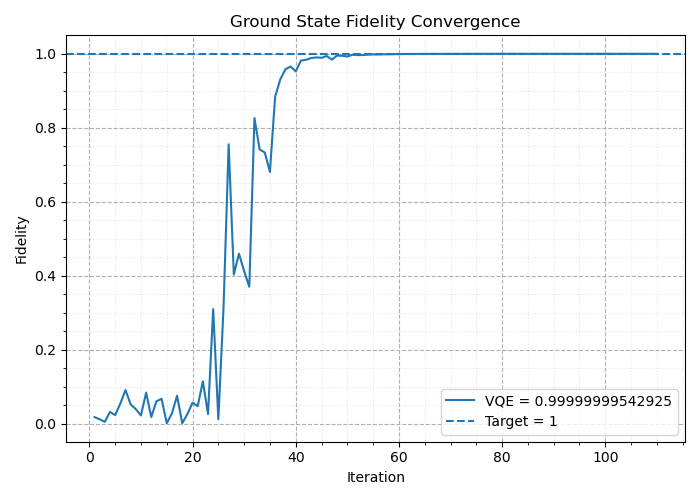}
    \caption{Ground-state fidelity convergence for 8-qubit flavor-number-conserving ansatz with $m=0$, $g^2=1.0$, $\Nf=2$, $\LD=2$, and open boundary conditions.}
    \label{fig:gs_fidelity}
\end{figure}

The statevector and matrix-product-state circuit simulations reported below were performed classically using CUDA-Q~\cite{cudaq2023}; the two-flavor exact charge-sector calculation described in Sec.~\ref{sec:sc_adapt_stabilization} uses a separate implementation. For the fixed circuit ansatze, we use SciPy's Sequential Least Squares Programming (SLSQP) optimizer with a tolerance of $10^{-10}$ and a maximum of 1000 iterations. The examples in Figs.~\ref{fig:gs_energy} and~\ref{fig:gs_fidelity} converge before this limit. The adaptive construction also utilizes the SLSQP optimizer, but with a tolerance of $1\times10^{-8}$ and a numerical-Jacobian finite-difference step size of $1\times10^{-4}$ for a maximum of $300$ iterations. For the additional one-flavor source calculations at $\LD=10$-14 used in the extrapolation of Sec.~\ref{sec:sc_adapt_stabilization}, the prescribed ten-generator architecture is reoptimized with SLSQP using $\texttt{ftol}=10^{-12}$, a finite-difference step of $10^{-6}$, and at most 300 iterations.

All calculations use open boundary conditions and lattice spacing $a=1$. Reference and variational energies use the same energy-zero prescription, including the interaction identity contribution. The ground-state reference energies are obtained with DMRG, with agreement with exact diagonalization to machine precision for the checked systems through 16 qubits. The two-flavor fixed results at $g^2=1.0$ as well as later benchmark points use five layers for both the total-number and flavor-number-conserving ans\"atze. Fixed circuits use convention $I$, while SC-ADAPT uses convention $R$, raw-score selection with $\delta=10^{-3}$, and the interaction-family normalization $\eta_{gh}=0.2$; the unitary equivalence of the two hopping conventions is discussed in Sec.~\ref{sec:jw_convention}.

We assess each variational result against the corresponding reference energy of the same finite-volume Hamiltonian. For $X\in\{N,F,\mathrm{SC}\}$, define the relative energy error
\begin{equation}
  \epsilon_{E,X}
  =
  \frac{|E_X-E_0|}{|E_0|},
  \label{eq:relative_energy_error}
\end{equation}
where $E_0$ is the ground-state reference energy; $E_N$ and $E_F$ are obtained with the total-number-conserving and flavor-number-conserving fixed ansatz families, respectively, all respecting the graph of the Hamiltonian; and $E_{\mathrm{SC}}$ is the SC-ADAPT result. The quantity $\epsilon_E$ is an energy diagnostic, not a state fidelity. The error values in the small-system benchmark tables are reconstructed from the displayed energies and inherit their rounding precision. In Table~\ref{tab:nf1_large_mps_transfer}, the percentage errors are instead calculated from the unrounded energies. There, MPS simulations evaluate the transferred circuits, while independently optimized DMRG states provide the ground-state references. The MPS calculations use CUDA-Q 0.14.2 in double precision with maximum bond dimension 256. The absolute and relative truncation cutoffs are both $10^{-13}$ for $\LD=30,50,100$ and both $10^{-12}$ for $\LD=200$, where the GESVD singular-value decomposition is used.

\subsection{Two-Flavor Comparison}
\label{sec:two_flavor_results}

For $\Nf=2$, the register contains $N_q=4\LD$ qubits. Tables~\ref{tab:massless_two_flavor} and~\ref{tab:massive_two_flavor} cover the common range $\LD=3$-$7$ for the fixed as well as SC-ADAPT constructions. For both massless and massive comparisons at $\LD=3$-$7$, the displayed errors satisfy $\epsilon_{E,F}<\epsilon_{E,N}<\epsilon_{E,\mathrm{SC}}$. For the fixed constructions, this ordering concerns circuits which are matched in optimization hyperparameters, circuit depth, and block count, thus isolating the charge preservation as its cause.

\begin{table}[t]
  \centering
  \small
  \caption{Ground-state energies for $m=0$, $g^2=1.0$, and $\Nf=2$, with $a=1$ and open boundary conditions.}
  \label{tab:massless_two_flavor}
  \setlength{\tabcolsep}{4pt}
  \renewcommand{\arraystretch}{1.15}
  \begin{tabular*}{\linewidth}{@{\extracolsep{\fill}}rrrrr@{}}
    \hline\hline
    \multicolumn{1}{c}{$\LD$} &
    \multicolumn{1}{c}{$E_0$} &
    \multicolumn{1}{c}{$E_N$} &
    \multicolumn{1}{c}{$E_F$} &
    \multicolumn{1}{c}{$E_{\mathrm{SC}}$} \\
    \hline
    3 & $-7.417$  & $-7.388$  & $-7.413$ & $-7.244$  \\
    4 & $-9.919$  & $-9.903$  & $-9.912$ & $-9.562$  \\
    5 & $-12.422$ & $-12.395$ & $-12.408$ & $-11.988$ \\
    6 & $-14.924$ & $-14.894$ & $-14.902$ & $-14.382$ \\
    7 & $-17.426$ & $-17.370$ & $-17.398$ & $-16.745$ \\
    \hline\hline
  \end{tabular*}
\end{table}

\begin{table}[t]
  \centering
  \small
  \caption{Ground-state energies for $m=0.5$, $g^2=1.0$, and $\Nf=2$, with $a=1$ and open boundary conditions.}
  \label{tab:massive_two_flavor}
  \setlength{\tabcolsep}{4pt}
  \renewcommand{\arraystretch}{1.15}
  \begin{tabular*}{\linewidth}{@{\extracolsep{\fill}}rrrrr@{}}
    \hline\hline
    \multicolumn{1}{c}{$\LD$} &
    \multicolumn{1}{c}{$E_0$} &
    \multicolumn{1}{c}{$E_N$} &
    \multicolumn{1}{c}{$E_F$} &
    \multicolumn{1}{c}{$E_{\mathrm{SC}}$} \\
    \hline
    3 & $-9.7414$  & $-9.7410$ & $-9.7411$ & $-9.7407$ \\
    4 & $-13.0276$ & $-13.0237$ & $-13.0275$ & $-13.0221$ \\
    5 & $-16.3140$ & $-16.3130$ & $-16.3137$ & $-16.2999$ \\
    6 & $-19.6002$ & $-19.5974$ & $-19.5999$  & $-19.5806$ \\
    7 & $-22.8865$ & $-22.8844$ & $-22.8861$ & $-22.8617$ \\
    \hline\hline
  \end{tabular*}
\end{table}

For each construction $X$, we note that $\epsilon_{E,X,m\ne0}\leq\epsilon_{E,X,m=0}$ at matched circuit depth and parameter count. While this isolates the non-zero mass as the source for the improved accuracy across $\LD=3$-$7$, this is simply an empirical observation for both the fixed and adaptive circuits. 

\subsection{Flavor-Number-Conserving Performance Across Flavor Number}
\label{sec:n_flavor_results}

Having empirically confirmed the improved performance with stronger symmetry constraints, the flavor-number-conserving ansatz is also tested across multiple flavor families and in a different coupling regime, benchmarking the accuracy against the configurations reported in Ref.~\cite{bakalov2026massiveGN} with $m=0.5$ and $g^2=0.4$, where available. As mentioned previously, the lattice Hamiltonian used by the authors of Ref.~\cite{bakalov2026massiveGN} uses a different fermion discretization, and both Hamiltonians thus have different spectra in the current finite limit. We thus compare only the relative energy errors reported in their work. The flavor-number-conserving ansatz improves upon reported benchmarks for the systems with up to 20 qubits tested in Ref.~\cite{bakalov2026massiveGN} and is further tested on systems through 30 qubits.

\begin{table}[t]
  \centering
  \small
  \caption{Ground-state energies obtained by the flavor-number-conserving ansatz for $m=0.5$ and $g^2=0.4$, with $a=1$ and open boundary conditions. The qubit count is $N_q=2\LD\Nf$. Relative energy errors $\epsilon_{E,F}$ are calculated from the rounded energies shown and compared with the errors $\bar\epsilon_{E}$ reported in Ref.~\cite{bakalov2026massiveGN} where they are available.}
  \label{tab:n_flavor}
  \setlength{\tabcolsep}{4pt}
  \renewcommand{\arraystretch}{1.15}
  \begin{tabular*}{\linewidth}{@{\extracolsep{\fill}}ccrrrr@{}}
    \hline\hline
    \multicolumn{1}{c}{$\Nf$} &
    \multicolumn{1}{c}{$N_q$} &
    \multicolumn{1}{c}{$E_0$} &
    \multicolumn{1}{c}{$E_{F}$} &
    \multicolumn{1}{c}{$\epsilon_{E,F}$} &
    \multicolumn{1}{c}{$\bar\epsilon_{E}$} \\
    \hline
    \multirow{8}{*}{1} & 16 & $-7.9873$  & $-7.9872$  & $1\times10^{-5}$ & $4\times10^{-4}$\\
     & 18 & $-8.9996$  & $-8.9993$  & $3\times10^{-5}$ & $3\times10^{-4}$\\
     & 20 & $-10.0119$ & $-10.0110$ & $9\times10^{-5}$ & $3\times10^{-4}$\\
     & 22 & $-11.0242$ & $-11.0229$ & $1\times10^{-4}$ & - \\
     & 24 & $-12.0365$ & $-12.0355$ & $1\times10^{-4}$ & - \\
     & 26 & $-13.0489$ & $-13.0481$ & $6\times10^{-5}$ & - \\
     & 28 & $-14.0612$ & $-14.0602$ & $7\times10^{-5}$ & - \\
     & 30 & $-15.0735$ & $-15.0716$ & $1\times10^{-4}$ & - \\
    \hline
    \multirow{4}{*}{2} & 16 & $-8.8232$  & $-8.8226$  & $7\times10^{-5}$ & $2\times10^{-4}$\\
     & 20 & $-11.0759$ & $-11.0749$ & $9\times10^{-5}$ & $5\times10^{-4}$\\
     & 24 & $-13.3285$ & $-13.3276$ & $7\times10^{-5}$ & - \\
     & 28 & $-15.5811$ & $-15.5804$ & $4\times10^{-5}$ & - \\
    \hline
    \multirow{2}{*}{3}  & 18 & $-11.2095$ & $-11.2091$ & $4\times10^{-5}$ & $2\times10^{-4}$\\
     & 24 & $-15.0222$ & $-15.0215$ & $5\times10^{-5}$ & - \\
    \hline
    \multirow{2}{*}{4} & 16 & $-11.2340$ & $-11.2336$ & $4\times10^{-5}$ & $5\times10^{-5}$\\
     & 24 & $-16.9764$ & $-16.9737$ & $2 \times10^{-4}$ & - \\
    \hline\hline
  \end{tabular*}
\end{table}

Table \ref{tab:n_flavor} shows that the flavor-number-conserving ansatz achieves a higher accuracy than all available current benchmarks $\bar\epsilon_{E}$. It is important to note that while all circuits are matched at depth, the parameter count is dependent on the number of flavors.

\subsection{SC-ADAPT Results Across Flavor Number}
\label{sec:sc_adapt_flavor_results}

Table~\ref{tab:sc_adapt_multiflavor} gives SC-ADAPT energies at $m=0.5$ and $g^2=0.4$ for $\Nf=1,2,3$. For the large-volume one-flavor benchmark, the volume family was extended to $d\leq9$; the default cap in Eq.~\eqref{eq:numerical_pool} was used otherwise. Across the selected systems with $\Nf=1,2,3$, which reach $N_q=30$, the relative energy error remains below $2.1\times10^{-3}$.

\begin{table}[t]
  \centering
  \footnotesize
  \caption{SC-ADAPT ground-state energies for $m=0.5$ and $g^2=0.4$, with $a=1$, open boundary conditions, and $\eta_{gh}=0.2$. The qubit count is $N_q=2\LD\Nf$. Relative energy errors are calculated from the rounded energies shown.}
  \label{tab:sc_adapt_multiflavor}
  \setlength{\tabcolsep}{4pt}
  \renewcommand{\arraystretch}{1.15}
  \begin{tabular*}{\linewidth}{@{\extracolsep{\fill}}ccrrr@{}}
    \hline\hline
    \multicolumn{1}{c}{$\Nf$} &
    \multicolumn{1}{c}{$\LD$} &
    \multicolumn{1}{c}{$E_0$} &
    \multicolumn{1}{c}{$E_{\mathrm{SC}}$} &
    \multicolumn{1}{c}{$\epsilon_{E,\mathrm{SC}}$} \\
    \hline
    \multirow{9}{*}{1} & 7  & $-6.9750$  & $-6.9642$  & $1.5\times10^{-3}$ \\
     & 8  & $-7.9873$  & $-7.9740$  & $1.7\times10^{-3}$ \\
     & 9  & $-8.9996$  & $-8.9838$  & $1.8\times10^{-3}$ \\
     & 10 & $-10.0119$ & $-9.9936$  & $1.8\times10^{-3}$ \\
     & 11 & $-11.0242$ & $-11.0035$ & $1.9\times10^{-3}$ \\
     & 12 & $-12.0365$ & $-12.0133$ & $1.9\times10^{-3}$ \\
     & 13 & $-13.0489$ & $-13.0232$ & $2.0\times10^{-3}$ \\
     & 14 & $-14.0612$ & $-14.0330$ & $2.0\times10^{-3}$ \\
     & 15 & $-15.0735$ & $-15.0429$ & $2.0\times10^{-3}$ \\
    \hline
    \multirow{5}{*}{2} & 3  & $-6.5707$  & $-6.5692$  & $2.3\times10^{-4}$ \\
     & 4  & $-8.8232$  & $-8.8190$  & $4.8\times10^{-4}$ \\
     & 5  & $-11.0759$ & $-11.0680$ & $7.1\times10^{-4}$ \\
     & 6  & $-13.3285$ & $-13.3176$ & $8.2\times10^{-4}$ \\
     & 7  & $-15.5811$ & $-15.5665$ & $9.4\times10^{-4}$ \\
    \hline
    \multirow{4}{*}{3} & 2  & $-7.3968$  & $-7.3954$  & $1.9\times10^{-4}$ \\
     & 3  & $-11.2095$ & $-11.2084$ & $9.8\times10^{-5}$ \\
     & 4  & $-15.0222$ & $-15.0183$ & $2.6\times10^{-4}$ \\
     & 5  & $-18.8350$ & $-18.8290$ & $3.2\times10^{-4}$ \\
    \hline\hline
  \end{tabular*}
\end{table}

\subsection{Sensitivity of SC-ADAPT Operator Selection}
\label{sec:sc_adapt_selection_audit}

We abbreviate $O_{mh}^{V,S0,S1}(d)$ as $V(d),S_0(d),S_1(d)$ and $O_{gh}^{V}(d)$ as $V_{gh}(d)$.

We examine the sensitivity of operator selection to the finite-difference
step and the relative normalization of the interaction-hopping family
for $\Nf=2$, $\LD=4$, $m=0.5$, and $g^2=0.4$, using the nine-element
pool in Eq.~\eqref{eq:numerical_pool}. Energy expectation values are
evaluated with noiseless statevectors in CUDA-Q 0.14.2. At each step,
all variational parameters are reoptimized using SLSQP with
$\texttt{maxiter}=300$, $\texttt{eps}=10^{-4}$, and
$\texttt{ftol}=10^{-8}$. For each of
$\eta_{gh}=0.1,0.2,0.3$, we perform seven append-and-reoptimization
steps; the sequence length is fixed independently of an adaptive
convergence criterion. All reoptimizations terminate successfully. The selection scores and optimized energies for $\eta_{gh}=0.2$ are listed in Table~\ref{tab:sc_adapt_selection_audit}
of Appendix~\ref{app:sc_adapt_details}.

At every one of the 21 pre-append states, the raw-score leader is unchanged over $\mathcal D=\{10^{-4},10^{-3},10^{-2}\}$. For the candidate $i_k$ selected at $\delta=10^{-3}$, define
\begin{equation}
  s_{\delta,k}
  =
  \frac{
  \max_{\delta\in\mathcal D}|g_{i_k}^{\mathrm{FD}}(\delta)|
  -\min_{\delta\in\mathcal D}|g_{i_k}^{\mathrm{FD}}(\delta)|
  }{|g_{i_k}^{\mathrm{FD}}(10^{-3})|}.
  \label{eq:delta_score_spread}
\end{equation}
The maximum of $s_{\delta,k}$ over the 21 decisions is approximately $1.1\times10^{-3}$. By contrast, the raw trajectory depends on $\eta_{gh}$: $\eta_{gh}=0.1$ and $0.2$ give the same sequence and final energies differing by $2.2\times10^{-7}$, while $\eta_{gh}=0.3$ selects $V_{gh}(1)$ first and follows a different sequence. At the reference state, the crossover of the two leading raw scores occurs near $\eta_{gh}=0.25$. Thus $\eta_{gh}$ is an explicit algorithmic hyperparameter of the raw-score prescription.

For each candidate, define the finite-step states $|\psi_i(\theta)\rangle=\widetilde U_i(\theta)|\psi\rangle$ at the current raw-selected state. As a normalization diagnostic, we evaluate the finite-step Fubini-Study tangent scale
\begin{equation}
  \sigma_i^{\mathrm{FS}}(\epsilon)
  =
  \frac{
    \sqrt{1-
    \bigl|\langle\psi_i(-\epsilon)|\psi_i(+\epsilon)\rangle\bigr|^2}
  }{2\epsilon}
  \label{eq:fs_tangent_scale}
\end{equation}
and the ratio $|g_i^{\mathrm{FD}}|/\sigma_i^{\mathrm{FS}}$ along the raw-selected trajectories, using $\epsilon=10^{-3}$ and checking $\epsilon=10^{-4}$-$10^{-2}$. At the reference state, this diagnostic gives approximately $3.7417$ for both $V(1)$ and $V_{gh}(1)$, a near tie at the resolution of the finite-step estimator. At steps 2-6 of the $\eta_{gh}=0.2$ raw trajectory, the diagnostic and raw score have the same leader, whereas the diagnostic favors $V_{gh}(3)$ rather than $V(5)$ at step 7. These values are evaluated on states generated by raw selection and do not constitute a normalized-selection trajectory.

\subsection{Finite-Volume Patterns and One-Flavor Parameter Transfer}
\label{sec:sc_adapt_stabilization}

Finite-volume sequence patterns provide the diagnostic used for circuit transfer in the Schwinger-model SC-ADAPT construction~\cite{PRXQuantum.5.020315}. Table~\ref{tab:sc_adapt_operator_stabilization} lists the selected operator sequences. For one flavor, $V_{gh}$ is absent and $\eta_{gh}$ is irrelevant.

\begin{table}[t]
  \centering
  \footnotesize
  \caption{SC-ADAPT generator sequences for $m=0.5$, $g^2=0.4$, and $\eta_{gh}=0.2$. The one-flavor rows give complete selected sequences. For two flavors, the $\LD=4$ row gives the seven-append sequence at $\eta_{gh}=0.2$. The $\LD=5-7$ entries permit comparison of the displayed operator sequences, but complete step histories under a common stopping criterion are unavailable.}
  \label{tab:sc_adapt_operator_stabilization}
  \setlength{\tabcolsep}{4pt}
  \renewcommand{\arraystretch}{1.15}
  \begin{tabular*}{\linewidth}{@{\extracolsep{\fill}}ccp{0.72\linewidth}@{}}
    \hline\hline
    \multicolumn{1}{c}{$\Nf$} &
    \multicolumn{1}{c}{$\LD$} &
    Sequence in append order \\
    \hline
    \multirow{6}{*}{1} & 7 &
    $V(1)$, $V(3)$, $V(5)$, $S_0(1)$, $S_1(1)$, $S_1(3)$, $S_0(3)$, $V(7)$, $V(1)$ \\
    \addlinespace[4pt]
     & 8-11 &
    $V(1)$, $V(3)$, $V(5)$, $S_0(1)$, $S_1(1)$, $V(7)$, $S_1(3)$, $S_0(3)$, $V(1)$, $V(9)$ \\
    \addlinespace[4pt]
     & 12-15 &
    $V(1)$, $V(3)$, $V(5)$, $S_0(1)$, $V(7)$, $S_1(1)$, $S_1(3)$, $S_0(3)$, $V(1)$, $V(9)$ \\
    \hline
    \multirow{4}{*}{2} & 4 &
    $V(1)$, $V(3)$, $V_{gh}(1)$, $S_1(1)$, $S_0(1)$, $S_1(3)$, $V(5)$ \\
    \addlinespace[4pt]
     & 5-6 &
    $V(1)$, $V(3)$, $V_{gh}(1)$, $S_0(1)$ \\
    \addlinespace[4pt]
     & 7 &
    $V(1)$, $V(3)$, $V_{gh}(1)$, $S_0(1)$, $V(5)$, $S_1(1)$, $V_{gh}(3)$ \\
    \hline\hline
  \end{tabular*}
\end{table}

For $\Nf=1$, the first four generators are common throughout $\LD=7$-15, while the complete ten-generator sequence is unchanged only for $\LD=12$-15. We therefore performed a retrospective transfer test from $\LD=12$ to $\LD=13,14,15$. The ten-generator append sequence and its tabulated four-decimal $\LD=12$ coordinates were held fixed. At each target volume, the translated volume factors were extended over the target bulk and the surface factors were reconstructed at the new endpoints. No target-side operator selection or parameter optimization was performed. The evaluations use noiseless statevectors in CUDA-Q 0.14.2 with double precision. The frozen coordinates are given in Appendix~\ref{app:sc_adapt_details}.

For the following comparison, define
\begin{equation}
  \epsilon_{E,\mathrm{eval}}
  =\frac{|E_{\mathrm{eval}}-E_0|}{|E_0|}.
  \label{eq:transfer_energy_error}
\end{equation}

\begin{table}[t]
  \centering
  \footnotesize
  \caption{One-flavor finite-volume transfer at $m=0.5$, $g^2=0.4$, $a=1$, and open boundaries. The $\LD=12$ row is the evaluation at the source volume, and the remaining rows use its frozen ten-generator sequence and ten tabulated four-decimal coordinates. $N_{R_\pm}$ is the total number of variable-angle current $R_-$ primitives and fixed-angle X-pattern conjugation $R_+$ primitives before compilation. The additional digits in $E_{\mathrm{eval}} $ resolve evaluations of the specified four-decimal parameters; they do not imply additional precision in the optimized parameters.}
  \label{tab:nf1_sc_adapt_transfer}
  \setlength{\tabcolsep}{4pt}
  \renewcommand{\arraystretch}{1.15}
  \begin{tabular*}{\linewidth}{@{\extracolsep{\fill}}ccccc@{}}
    \hline\hline
    \multicolumn{1}{c}{$\LD$} &
    \multicolumn{1}{c}{$N_q$} &
    \multicolumn{1}{c}{$N_{R_\pm}$} &
    \multicolumn{1}{c}{$E_{\mathrm{eval}}$} &
    \multicolumn{1}{c}{$10^3\epsilon_{E,\mathrm{eval}}$} \\
    \hline
    12 & 24 & 822  & $-12.013326$ & 1.925 \\
    \hline
    13 & 26 & 914  & $-13.023173$ & 1.972 \\
    14 & 28 & 1006 & $-14.033021$ & 2.004 \\
    15 & 30 & 1098 & $-15.042869$ & 2.032 \\
    \hline\hline
  \end{tabular*}
\end{table}

For the frozen ten-generator one-flavor sequence, the translated factors give
\begin{equation}
  \begin{aligned}
    N_{R_\pm}&=92\LD-282,\\
    N_{\mathrm{par}}&=10.
  \end{aligned}
  \label{eq:nf1_transfer_resource_count}
\end{equation}
The counts in Table~\ref{tab:nf1_sc_adapt_transfer} thus realize the scalable construction over $\LD=12$-15: the shared-coordinate count is fixed while the precompilation primitive count grows linearly with volume. These counts refer to the ansatz factors and do not include reference preparation, measurements, routing, or compiler cancellations; they are not compiled circuit depths.

The relative energy error changes only from $1.925\times10^{-3}$ in the source-volume evaluation to $2.032\times10^{-3}$ at $\LD=15$. For comparison, let $E_{\mathrm{opt}}^{\mathrm{replay}}$ denote the energy evaluated at the independently optimized rounded target coordinates and define the absolute energy difference $\Delta E=E_{\mathrm{eval}}-E_{\mathrm{opt}}^{\mathrm{replay}}$. Evaluating those coordinates lowers the $\LD=13,14,15$ energies by only $1.41\times10^{-6}$, $6.11\times10^{-6}$, and $1.20\times10^{-5}$, respectively, in lattice energy units. These differences compare evaluations of rounded coordinate vectors and are not full-precision optimization gaps.

The calculation demonstrates finite-volume one-flavor parameter transfer within this scalable circuit construction: the frozen shared coordinates retain comparable energy accuracy over the tested window. The stable interval was identified retrospectively and the source coordinates and reference energies are rounded, so the demonstrated accuracy concerns this finite-volume energy test. State fidelity, local observables, and transfer of the multi-flavor $V_{gh}$ family have not been tested. Neither fixed-accuracy performance at arbitrary volumes nor efficient end-to-end classical simulation and optimization follows from the linear circuit-factor count.

We extend the one-flavor evaluations to $\LD=30,50,100,200$ at the same $m=0.5$, $g^2=0.4$, and half filling. The first three volumes use the frozen four-decimal $\LD=12$ parameters. For $\LD=200$, the common ten-generator architecture shown in the $\LD=12$-15 row of Table~\ref{tab:sc_adapt_operator_stabilization} is first reoptimized at each source volume $\LD=10$-14. These calculations impose the same ordered architecture at every source volume; they do not repeat adaptive operator selection. The resulting coordinates are fitted separately to
\begin{equation}
  \theta_i(\LD)=\theta_{i,\infty}+C_i e^{-b_i(\LD-10)},
  \label{eq:nf1_parameter_extrapolation}
\end{equation}
with $0.01\leq b_i\leq2$. The empirical coordinate-resolution criterion is $10^{-6}$: a retained fit must resolve the finite-volume variation and have a maximum absolute residual no larger than this value. Coordinates 3, 5, and 8 in the stated append order satisfy these criteria and are evaluated at $\LD=200$; the other seven retain their optimized $\LD=14$ values. This prescription does not establish exponential convergence of all ten coordinates, and the resolution criterion is not a statistical uncertainty. No target-side operator selection or parameter optimization is performed.

\begin{table}[t]
  \centering
  \footnotesize
  \caption{One-flavor circuit evaluations at $m=0.5$, $g^2=0.4$, and $a=1$, with open boundaries and half filling. Energy densities are $\mathcal{E}=E/\LD$; percentage errors are measured relative to the corresponding DMRG reference. All MPS evaluations use maximum bond dimension 256. The first three rows use frozen $\LD=12$ parameters; the final row uses the extrapolation prescription described in the text.}
  \label{tab:nf1_large_mps_transfer}
  \setlength{\tabcolsep}{4pt}
  \renewcommand{\arraystretch}{1.15}
  \begin{tabular*}{\linewidth}{@{\extracolsep{\fill}}ccccc@{}}
    \hline\hline
    \multicolumn{1}{c}{$\LD$} &
    \multicolumn{1}{c}{Qubits} &
    \multicolumn{1}{c}{$\mathcal{E}_{\text{DMRG}}$} &
    \multicolumn{1}{c}{$\mathcal{E}_{\text{MPS}}$} &
    \multicolumn{1}{c}{$\epsilon_{\text{MPS}} (\%)$} \\
    \hline
    30  & 60  & $-1.008607$ & $-1.005082$ & 0.3495 \\
    50  & 100 & $-1.010090$ & $-1.004836$ & 0.5202 \\
    100 & 200 & $-1.011203$ & $-1.001237$ & 0.9856 \\
    200 & 400 & $-1.011759$ & $-0.993962$ & 1.7590 \\
    \hline\hline
  \end{tabular*}
\end{table}

Table~\ref{tab:nf1_large_mps_transfer} gives the MPS energy densities $\mathcal{E}=E/\LD$ and $100\epsilon_{E,\mathrm{eval}}$, with $E_{\mathrm{eval}}=E_{\mathrm{MPS}}$ and $E_0=E_{\mathrm{DMRG}}$ in Eq.~\eqref{eq:transfer_energy_error}. The deviations from DMRG are approximately $0.35\%$-$1.76\%$ across these 60-400-qubit evaluations. The finite-bond MPS evaluations retain appreciable simulation error, so these deviations cannot be attributed solely to the variational circuits. The different parameter prescriptions and numerical settings also preclude interpreting the rows as a controlled scaling law for the energy error.

For $N_f=2$, the leading sequence $V(1),V(3),V_{gh}(1)$ recurs over $L_D=4$–$7$, although the available results do not yet establish stabilization of the full operator sequence. Keeping the seven-generator sequence obtained at $L_D=7$ fixed, we successfully optimized its parameters at $L_D=8$ within the joint charge sector $N_0=N_1=8$. Extending these calculations to larger volumes is limited by the cost and accuracy of classical circuit evaluation. The statevector in the fixed charge sector still requires exponentially growing memory, while the MPS calculations at the tested bond dimensions retain appreciable discrepancies from the exact energy of the same $L_D=8$ circuit. We therefore restrict the large-volume transfer tests to the one-flavor case.

\section{Excited States}
\label{sec:excited_states}

The flavor-number-conserving fixed ansatz can also be used to target successive states in a chosen joint charge sector. We define the normalized trial state
\(
|\psi(\boldsymbol\theta)\rangle
=U(\boldsymbol\theta)|\psi_{\mathrm{ref}}\rangle
\)
and minimize the sequentially deflated objective
\begin{equation}
  \begin{aligned}
  \mathcal L_k(\boldsymbol\theta)
  ={}&\langle\psi(\boldsymbol\theta)|H|\psi(\boldsymbol\theta)\rangle 
  +\sum_{i=0}^{k-1}\lambda_i
  \left|\langle\widetilde\Psi_i|\psi(\boldsymbol\theta)\rangle\right|^2,
  \end{aligned}
  \label{eq:general_excited_loss}
\end{equation}
where $k\geq1$ is the zero-based rank within the chosen sector and $|\widetilde\Psi_i\rangle$ are normalized VQE states prepared at the preceding ranks in that sector. The ground-state objective is $\mathcal L_0=\langle\psi|H|\psi\rangle$. For the massless and massive benchmarks, we set $\lambda_i=\lambda=10$ in lattice energy units. Because the Hamiltonian commutes with each flavor charge, its complete finite-volume spectrum decomposes into distinct joint charge sectors. A flavor-number-conserving circuit initialized in $(N_0,N_1)=(2,2)$ restricts the variational search to that sector, as in charge-preserving deflation methods \cite{Chandani:2024xza}. Exact charge conservation fixes the search sector but does not by itself guarantee convergence or completeness at finite circuit depth.

For a nondegenerate exact target $|\Psi_k\rangle$, we report the bare variational energy, rather than the penalized objective, through
\begin{equation}
  \begin{aligned}
  E_{\mathrm{VQE}}^{(k)}&=\langle\psi_k|H|\psi_k\rangle, \\
  \epsilon_E^{(k)}&=
  \frac{|E_{\mathrm{VQE}}^{(k)}-E_{\mathrm{exact}}^{(k)}|}
       {|E_{\mathrm{exact}}^{(k)}|}, \\
  \epsilon_F^{(k)}&=1-|\langle\Psi_k|\psi_k\rangle|^2.
  \end{aligned}
  \label{eq:excited_state_diagnostics}
\end{equation}

We characterize the charge sectors of the four targeted levels and compare the prepared states $|\psi_i\rangle$ with the exact eigenstates $|\Psi_i\rangle$. Table~\ref{tab:excited_energies} gives the relative energy errors. The squared overlaps with the exact eigenstates and between the prepared states have the ideal values

\begin{equation}
    \begin{aligned}
        |\langle\Psi_i|\psi_j\rangle|^2&=\delta_{ij}\\
        |\langle\psi_i|\psi_j\rangle|^2&=\delta_{ij},
    \end{aligned}
    \label{eq:orthogonality}
\end{equation}

with the latter condition being a self-consistency check for the VQE, as both squared overlaps have the identity matrix as an ideal target, and approximate agreement is achieved for the VQE states. 

\begin{table}[t]
  \centering
  \footnotesize
  \caption{Energy discrepancies for the lowest four states in the $(N_0,N_1)=(2,2)$ sector at $g^2=1.0$, $a=1$, $\LD=2$, and $\Nf=2$ with open boundary conditions for $m=0$ and $m=0.5$. Within-sector ranks $k=0,1,2,3$ correspond to global spectrum indices 0, 1, 2, and 3 for the massless case and 0, 5, 6, and 14 for the massive case, respectively.}
  \label{tab:excited_energies}
  \setlength{\tabcolsep}{4pt}
  \renewcommand{\arraystretch}{1.15}
  \begin{tabular*}{\linewidth}{@{\extracolsep{\fill}}ccc@{}}
    \hline\hline
    Within-sector rank $k$ & $\epsilon_{E,m=0}$ & $\epsilon_{E,m=0.5}$ \\
    \hline
    0 & $2.93\times10^{-9}$ & $2.68\times10^{-9}$ \\
    1 & $2.06\times10^{-11}$ & $1.07\times10^{-11}$ \\
    2 & $6.47\times10^{-8}$ & $8.78\times10^{-7}$\\
    3 &  $1.20\times10^{-6}$ & $2.28\times10^{-8}$\\
    \hline\hline
  \end{tabular*}
\end{table}

\subsection{Extended Spectrum for Massless Hamiltonian}
\label{sec:massless_spectrum}

\begin{figure}[t]
    \centering        \includegraphics[width=\linewidth]{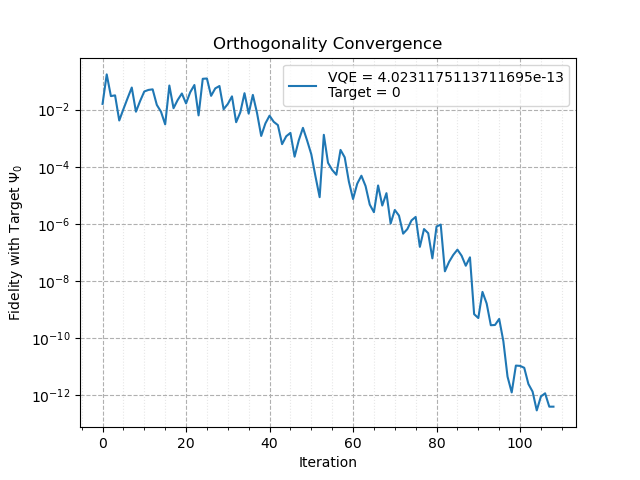}
    \caption{Squared overlap of VQE first excited state with the target VQE ground state minimized via the loss function with $m=0$, $g^2=1.0$, $a=1$, $\LD=2$, $\Nf=2$, open boundary conditions, and $\lambda=10$ applied to the deflation terms.}
    \label{fig:1excited_fidelity_gs}
\end{figure}

\begin{figure}[!htbp]
    \centering
    \includegraphics[width=\linewidth]{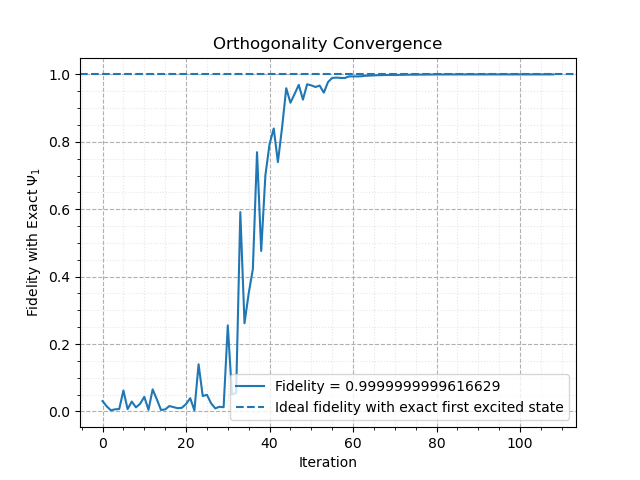}
    \caption{Squared overlap of VQE first excited state with the exact first excited state with $m=0$, $g^2=1.0$, $a=1$, $\LD=2$, $\Nf=2$, open boundary conditions, and $\lambda=10$ applied to the deflation terms.}
    \label{fig:1excited_fidelity_fes}
\end{figure}

\begin{figure}[t]
    \centering
    \includegraphics[width=\linewidth]{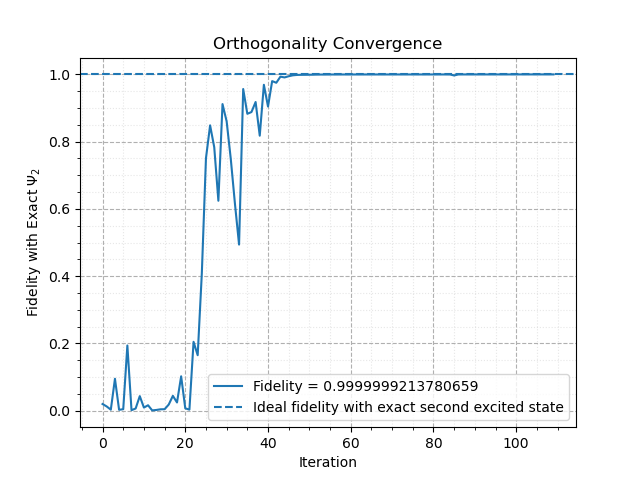}

    \caption{Squared overlap of VQE second excited state with the exact second excited state with $m=0$, $g^2=1.0$, $a=1$, $\LD=2$, $\Nf=2$, open boundary conditions, and $\lambda=10$ applied to the deflation terms.}
    \label{fig:2excited_second_fidelity}
\end{figure}

\begin{figure}
    \centering
    \includegraphics[width=\linewidth]{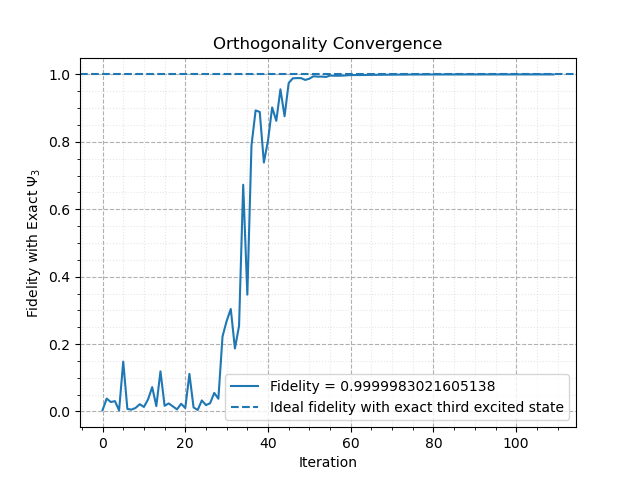}

    \caption{Squared overlap of the VQE third excited state with the exact third excited state with $m=0$, $g^2=1.0$, $a=1$, $\LD=2$, $\Nf=2$, open boundary conditions, and $\lambda=10$ applied to the deflation terms.}
    \label{fig:3excited_third_fidelity}
\end{figure}

For the massless Hamiltonian with $m=0$ and $g^2=1.0$, the first three excited states in the complete spectrum lie in the same number and flavor sector as the ground state. They are therefore not excluded by flavor-charge conservation and were targeted from the same reference. Figs. ~\ref{fig:1excited_fidelity_gs} and ~\ref{fig:1excited_fidelity_fes} show the overlap histories of the VQE first excited state with the orthogonal VQE ground state used in the loss function penalty, and the target state, respectively, while Figs. \ref{fig:2excited_second_fidelity} and \ref{fig:3excited_third_fidelity} show the corresponding overlap histories for the second and third excited states, respectively. Table \ref{tab:massless_fidelity_matrix_true} shows the fidelities between the VQE and target exact eigenstates, while table \ref{tab:massless_fidelity_matrix_vqe} shows the fidelities between the different VQE states.

\begin{table*}[t]
  \centering
  \small
  \caption{Fidelities between the VQE states $\psi$ and the exact eigenstates $\Psi$ of the $(N_0,N_1)=(2,2)$ sector of the massless eight-qubit benchmark with $m=0$, $g^2=1.0$, $a=1$, $\LD=2$, $\Nf=2$, open boundary conditions, and $\lambda=10$ applied to the deflation terms. The ideal case is an identity matrix.}
  \label{tab:massless_fidelity_matrix_true}
  \setlength{\tabcolsep}{4pt}
  \renewcommand{\arraystretch}{1.15}
  \begin{tabular*}{\linewidth}{@{\extracolsep{\fill}}lcccc@{}}
    \hline\hline
     &
    \multicolumn{1}{c}{$\ket{\psi_0}$} &
    \multicolumn{1}{c}{$\ket{\psi_1}$} &
    \multicolumn{1}{c}{$\ket{\psi_2}$} &
    \multicolumn{1}{c}{$\ket{\psi_3}$} \\
    \hline
    $\bra{\Psi_0}$ & $0.99999999543$ & $3.40\times10^{-12}$ & $3.57\times10^{-11}$ & $1.89\times10^{-11}$\\
    $\bra{\Psi_1}$ & $4.21\times10^{-12}$ & $0.99999999996$ & $5.86\times10^{-10}$ & $1.93\times10^{-12}$\\
    $\bra{\Psi_2}$ & $1.73\times10^{-11}$ & $4.78\times10^{-12}$ & $0.99999992138$ & $6.34\times10^{-10}$\\
    $\bra{\Psi_3}$ & $3.99\times10^{-12}$ & $8.84\times10^{-13}$ & $1.25\times10^{-11}$ & $0.99999830216$ \\
    \hline\hline
  \end{tabular*}
\end{table*}

\begin{table*}[t]
  \centering
  \small
  \caption{Fidelities between the VQE states $\psi$ of the $(N_0,N_1)=(2,2)$ sector of the massless eight-qubit benchmark with $m=0$, $g^2=1.0$, $a=1$, $\LD=2$, $\Nf=2$, open boundary conditions, and $\lambda=10$ applied to the deflation terms. The ideal case is an identity matrix.}
  \label{tab:massless_fidelity_matrix_vqe}
  \setlength{\tabcolsep}{4pt}
  \renewcommand{\arraystretch}{1.15}
  \begin{tabular*}{\linewidth}{@{\extracolsep{\fill}}lcccc@{}}
    \hline\hline
     &
    \multicolumn{1}{c}{$\ket{\psi_0}$} &
    \multicolumn{1}{c}{$\ket{\psi_1}$} &
    \multicolumn{1}{c}{$\ket{\psi_2}$} &
    \multicolumn{1}{c}{$\ket{\psi_3}$} \\
    \hline
    $\bra{\psi_0}$ & $1.00$ & $4.02\times10^{-13}$ & $1.33\times10^{-11}$ & $8.37\times10^{-12}$ \\
    $\bra{\psi_1}$ & $4.02\times10^{-13}$ & $1.00$ & $4.85\times10^{-10}$ & $6.94\times10^{-13}$ \\
    $\bra{\psi_2}$ & $1.33\times10^{-11}$ & $4.85\times10^{-10}$ & $1.00$ & $7.97\times10^{-10}$ \\
    $\bra{\psi_3}$ & $8.37\times10^{-12}$ & $6.94\times10^{-13}$ & $7.97\times10^{-10}$ & $1.00$ \\
    \hline\hline
  \end{tabular*}
\end{table*}

Table~\ref{tab:massless_spectrum_flavors} places these states in the complete energy-ordered spectrum. The next full-spectrum levels lie outside the $(2,2)$ reference sector and therefore cannot be reached from the same initialization by this flavor-number-conserving circuit.

\begin{table}[t]
  \centering
  \footnotesize
  \caption{Selected exact energies and charge sectors for the massless eight-qubit benchmark with $m=0$, $g^2=1.0$, $a=1$, $\LD=2$, $\Nf=2$, and open boundary conditions. Global indices are zero-based positions in the complete energy-ordered spectrum.}
  \label{tab:massless_spectrum_flavors}
  \setlength{\tabcolsep}{4pt}
  \renewcommand{\arraystretch}{1.15}
  \begin{tabular*}{\linewidth}{@{\extracolsep{\fill}}crccc@{}}
    \hline\hline
    \multicolumn{1}{c}{Global index} &
    \multicolumn{1}{c}{$E_{\mathrm{exact}}$} &
    \multicolumn{1}{c}{$N$} &
    \multicolumn{1}{c}{$N_0$} &
    \multicolumn{1}{c}{$N_1$} \\
    \hline
    0 & -4.9148 & 4 & 2 & 2 \\
    1 & -4.5697 & 4 & 2 & 2 \\
    2 & -4.4251 & 4 & 2 & 2 \\
    3 & -4.0811 & 4 & 2 & 2 \\
    \hline\hline
  \end{tabular*}
\end{table}

\subsection{Probing Particular Flavor Sectors for Massive Hamiltonian}
\label{sec:massive_flavors}

For the massive benchmark, the exact ground state lies in the joint charge sector $(N_0,N_1)=(2,2)$. As shown in Table~\ref{tab:massive_spectrum_flavors}, the four globally lowest excited states share a degenerate eigenvalue and lie in the distinct sectors $(3,2)$, $(2,3)$, $(1,2)$, and $(2,1)$. These are valid eigenstates of the Hamiltonian, but they cannot be reached from the $(2,2)$ reference by a circuit that preserves both flavor charges. The lowest exact excited eigenstate in the reference sector is therefore global index 5, which has within-sector rank $k=1$; global index 6 has within-sector rank $k=2$.

\begin{table}[t]
  \centering
  \footnotesize
  \caption{Selected exact energy levels for $m=0.5$, $g^2=1.0$, $a=1$, $\LD=2$, and $\Nf=2$ with open boundary conditions. Global indices are zero-based positions in the complete energy-ordered spectrum.}
  \label{tab:massive_spectrum_flavors}
  \setlength{\tabcolsep}{4pt}
  \renewcommand{\arraystretch}{1.15}
  \begin{tabular*}{\linewidth}{@{\extracolsep{\fill}}crcc@{}}
    \hline\hline
    \multicolumn{1}{c}{\shortstack{Global\\indices}} &
    \multicolumn{1}{c}{$E_{\mathrm{exact}}$} &
    \multicolumn{1}{c}{$N$} &
    \multicolumn{1}{c}{$(N_0,N_1)$} \\
    \hline
    0      & $-6.4552$ & 4 & $(2,2)$ \\
    1      & $-4.5813$ & 5 & $(3,2)$ \\
    2      & $-4.5813$ & 5 & $(2,3)$ \\
    3      & $-4.5813$ & 3 & $(1,2)$ \\
    4      & $-4.5813$ & 3 & $(2,1)$ \\
    5      & $-4.5072$ & 4 & $(2,2)$ \\
    6      & $-4.4968$ & 4 & $(2,2)$ \\
    13-15 & $-3.3390$ & 4 & $\{(1,3),(2,2),(3,1)\}$ \\
    \hline\hline
  \end{tabular*}
\end{table}

The level at $E_{\mathrm{exact}}=-3.3390$, represented by global indices 13-15, is an exactly degenerate flavor-$SU(2)$ triplet. An eigensolver that diagonalizes only $H$ may return an arbitrary orthonormal basis in this subspace, producing non-integer expectation values of $N_0$ and $N_1$ even though both charges are exactly conserved. Diagonalizing either charge within the degenerate subspace yields the simultaneous charge eigenstates $(1,3)$, $(2,2)$, and $(3,1)$. The $(2,2)$ member has within-sector rank $k=3$. Accordingly, a variational state should be compared either with that charge-resolved eigenvector or with the projector onto the complete degenerate subspace; squared overlaps with individual unrotated eigenvectors are basis-dependent diagnostics. For a degenerate target subspace $\mathcal D$, the basis-independent fidelity is instead
\begin{equation}
  \begin{aligned}
  F_{\mathcal D}&=\langle\psi_k|P_{\mathcal D}|\psi_k\rangle 
  =\sum_{\alpha\in\mathcal D}|\langle E_\alpha|\psi_k\rangle|^2, 
  \end{aligned}
  \label{eq:degenerate_projector_fidelity}
\end{equation}

\begin{table}[t]
  \centering
  \footnotesize
  \caption{Fidelities of the VQE third excited state lying in the $(2,2)$ sector with the triply degenerate flavor-diagonal states, as well as the degenerate subspace fidelity $F_D$ with the flavor-unresolved basis from \eqref{eq:degenerate_projector_fidelity} for $m=0.5$, $g^2=1.0$, $a=1$, $\LD=2$, and $\Nf=2$ with open boundary conditions. Within-sector rank $k=3$ corresponds to a triply degenerate subspace spanning global spectrum indices 13, 14, and 15.}
  \label{tab:deg_fidelity}
  \setlength{\tabcolsep}{4pt}
  \renewcommand{\arraystretch}{1.15}
  \begin{tabular*}{0.5\linewidth}{cc}
    \hline\hline
    \multicolumn{1}{c}{Quantity} &
    \multicolumn{1}{c}{Fidelity} \\
    \hline
    $F_{2,2}$ & $0.9999999462883007$ \\
    $F_{1,3}$ & $5.17\times10^{-31}$ \\
    $F_{3,1}$ & $2.71\times10^{-31}$ \\
    $F_D$ & $0.9999999462883011$ \\
    \hline\hline
  \end{tabular*}
\end{table}

While the fidelities of the VQE states with states lying in other flavor or number sectors are many orders of magnitude lower than with other orthogonal states in the same sector, both of these diagnostics - the fidelity with the exact charge-resolved $(2,2)$ state in a basis commuting with the flavor operator, as well as the basis-independent degenerate subspace fidelity give equivalent measures of circuit performance and are numerically close, as seen in Table \ref{tab:deg_fidelity}. The fidelities reported in all other tables, however, are those with the exact flavor-diagonal state. The penalty construction, therefore, targets successive variational states within a chosen flavor-number sector, while exact diagonalization organizes the complete spectrum across all sectors.

\begin{table*}[t]
  \centering
  \small
  \caption{Fidelities between the VQE states $\psi$ and the exact eigenstates $\Psi$ of the $(N_0,N_1)=(2,2)$ sector of the massive eight-qubit benchmark with $m=0.5$, $g^2=1.0$, $a=1$, $\LD=2$, $\Nf=2$, open boundary conditions, and $\lambda=10$ applied to the deflation terms. The ideal case is an identity matrix.}
  \label{tab:massive_fidelity_matrix_true}
  \setlength{\tabcolsep}{4pt}
  \renewcommand{\arraystretch}{1.15}
  \begin{tabular*}{\linewidth}{@{\extracolsep{\fill}}lcccc@{}}
    \hline\hline
     &
    \multicolumn{1}{c}{$\ket{\psi_0}$} &
    \multicolumn{1}{c}{$\ket{\psi_1}$} &
    \multicolumn{1}{c}{$\ket{\psi_2}$} &
    \multicolumn{1}{c}{$\ket{\psi_3}$} \\
    \hline
    $\bra{\Psi_0}$ & $ 0.99999999598$ & $3.07\times10^{-12}$ & $2.53\times10^{-9}$ & $1.41\times10^{-11}$\\
    $\bra{\Psi_1}$ & $2.49\times10^{-12}$ & $0.99999999998$ & $1.26\times10^{-8}$ & $5.73\times10^{-13}$\\
    $\bra{\Psi_2}$ & $4.49\times10^{-10}$ & $1.44\times10^{-13}$ & $0.99999859832$ & $7.73\times10^{-9}$\\
    $\bra{\Psi_3}$ & $6.72\times10^{-12}$ & $5.22\times10^{-13}$ & $3.00\times10^{-8}$ & $0.99999994629$ \\
    \hline\hline
  \end{tabular*}
\end{table*}

\begin{table*}[t]
  \centering
  \small
  \caption{Fidelities between the VQE states $\psi$ of the $(N_0,N_1)=(2,2)$ sector of the massive eight-qubit benchmark with $m=0.5$, $g^2=1.0$, $a=1$, $\LD=2$, $\Nf=2$, open boundary conditions, and $\lambda=10$ applied to the deflation terms. The ideal case is an identity matrix.}
  \label{tab:massive_fidelity_matrix_vqe}
  \setlength{\tabcolsep}{4pt}
  \renewcommand{\arraystretch}{1.15}
  \begin{tabular*}{\linewidth}{@{\extracolsep{\fill}}lcccc@{}}
    \hline\hline
     &
    \multicolumn{1}{c}{$\ket{\psi_0}$} &
    \multicolumn{1}{c}{$\ket{\psi_1}$} &
    \multicolumn{1}{c}{$\ket{\psi_2}$} &
    \multicolumn{1}{c}{$\ket{\psi_3}$} \\
    \hline
    $\bra{\psi_0}$ & $1.00$ & $3.06\times10^{-14}$ & $1.03\times10^{-9}$ & $1.40\times10^{-12}$ \\
    $\bra{\psi_1}$ & $3.06\times10^{-14}$ & $1.00$ & $1.26\times10^{-8}$ & $1.17\times10^{-13}$ \\
    $\bra{\psi_2}$ & $1.03\times10^{-9}$ & $1.26\times10^{-8}$ & $1.00$ & $7.31\times10^{-9}$ \\
    $\bra{\psi_3}$ & $1.40\times10^{-12}$ & $1.17\times10^{-13}$ & $7.31\times10^{-9}$ & $1.00$ \\
    \hline\hline
  \end{tabular*}
\end{table*}

\section{Conclusions and Outlook}
\label{sec:conclusions}

We have constructed and benchmarked symmetry-preserving fixed circuits and a problem-specific SC-ADAPT ansatz for ground-state and sector-resolved excited-state preparation in the lattice \GN{} model. The fixed families preserve either total fermion number or every flavor number. Their common circuit graph, layer organization, and local-phase parametrization permit individual flavor-number conservation to be imposed by removing cross-flavor exchange parameters. in the two-flavor benchmarks, this restriction improves the achieved energies under the reported optimization protocol, demonstrating the practical benefit of incorporating the flavor charges into these circuits.

The selected massive flavor-number-conserving benchmarks span $\Nf=1,2,3,4$ and registers through 30 qubits, with relative energy errors of approximately $1\times 10^{-5}$-$2\times10^{-4}$. At the available comparison points, these errors are smaller than those reported in Ref.~\cite{bakalov2026massiveGN}, each normalized to the reference energy of its respective finite-lattice Hamiltonian. The different discretizations and computational resources make this a comparison of reported benchmark errors rather than a matched algorithmic ranking.

The Gross-Neveu SC-ADAPT pool combines translated mass-hopping bulk and boundary-sensitive generators with a cross-flavor interaction-hopping family. The resulting circuit construction has linear precompilation local-factor growth and a volume-independent shared-parameter count at fixed $\Nf$, range caps, and append count. Selected massive benchmarks with $\Nf=1,2,3$ have relative energy errors below $2.1\times10^{-3}$ through 30 qubits. Transferring the ten-generator sequence and four-decimal parameters from $\LD=12$ to $\LD=13-15$ gives relative energy errors of $1.97\times10^{-3}$-$2.03\times10^{-3}$ without further optimization. Evaluations using independently optimized, rounded target parameters give comparable energies. The one-flavor study further extends circuit evaluations to 400 qubits without target-side optimization. At maximum MPS bond dimension 256, the reported energies differ from DMRG references by $0.35\%$-$1.76\%$, including finite-bond evaluation error. These results supplement the nearby-volume test without establishing fixed-accuracy performance at arbitrary volumes. For two flavors, the common leading sequence does not yet establish stabilization of the full operator ordering. Fixed-sequence optimization is feasible at $\LD=8$, while larger-volume transfer remains limited by the accuracy and cost of classical circuit evaluation.

Charge-preserving sequential deflation prepares the lowest four states by within-sector rank in the $(N_0,N_1)=(2,2)$ sector of both massless and massive eight-qubit systems. The displayed results have small energy discrepancies and near-unit squared-overlap or eigenspace-projector fidelities. For the massive rank-$3$ target, the unique $(2,2)$ component of the triply degenerate global eigenspace has a charge-resolved fidelity equal to the basis-independent projector fidelity. Resolving the charges and degeneracies is therefore essential when identifying the prepared spectral states.

The present results concern state preparation on finite lattices at fixed spacing. Continuum extrapolation and renormalized mass-gap scaling are separate objectives, as illustrated by tensor-network studies of the model \cite{roose2021latticeGN}. State fidelities, local observables, and controlled multi-flavor circuit evaluations would further assess transfer beyond the demonstrated energy benchmarks. Noisy circuit simulations and small hardware implementations provide direct next tests of preparation accuracy after compilation and in the presence of device errors. The prepared states also offer starting points for future real-time and scattering calculations \cite{jordan2014fermionic,li2024scatteringGN}.

\acknowledgments
We thank Talal Chowdhury for useful discussion. 
VB is supported in part by the US DOE under Award No DE-SC0024673. KK is supported in part by the National Science Foundation through Grant PHY-2609759. 
KY is supported by the Brookhaven National Laboratory LDRD 24-061, 25-033, and the U.S. Department of Energy, Office of Science, Grants No. DE-SC0012704. 
This research used resources of the National Energy Research Scientific Computing Center, a DOE Office of Science User Facility supported by the Office of Science of the U.S. Department of Energy under Contract No. DE-AC02-05CH11231 using NERSC award HEP-ERCAP0034458.  Additionally, this work also used resources available through the National Research Platform (NRP) at the University of California, San Diego \cite{10.1145/3708035.3736060}. NRP has been developed, and is supported in part, by funding from National Science Foundation, from awards 1730158, 1540112, 1541349, 1826967, 2112167, 2100237, and 2120019, as well as additional funding from community partners.  

\appendix

\section{Staggered and Jordan-Wigner Algebra}
\label{app:staggered_jw}

\subsection{Staggered Map and Phase Redefinition}

The staggered discretization associates the two components of the Dirac field in cell $n$ with the even and odd staggered sites:
\begin{equation}
  \begin{pmatrix}
    \psi_{1,b}\\
    \psi_{2,b}
  \end{pmatrix}_{\!n}
  \longleftrightarrow
  \frac{1}{\sqrt a}
  \begin{pmatrix}
    \chi_{2n,b}\\
    \chi_{2n+1,b}
  \end{pmatrix}.
  \label{eq:staggered_mapping}
\end{equation}
For the rephasing $\chi_{j,b}=i^j c_{j,b}$, each open-chain link obeys
\begin{equation}
  \begin{aligned}
    \chi_{j,b}^{\dagger}\chi_{j+1,b}
    &=i\,c_{j,b}^{\dagger}c_{j+1,b},
    \\
    \chi_{j+1,b}^{\dagger}\chi_{j,b}
    &=-i\,c_{j+1,b}^{\dagger}c_{j,b},
    \\
    \chi_{j,b}^{\dagger}\chi_{j,b}
    &=c_{j,b}^{\dagger}c_{j,b}.
  \end{aligned}
\end{equation}
Consequently,
\begin{equation}
  \begin{split}
    &-\frac{i}{2a}
    \left(
      \chi_{j,b}^{\dagger}\chi_{j+1,b}
      -\chi_{j+1,b}^{\dagger}\chi_{j,b}
    \right)
    \\
    &\qquad=
    \frac{1}{2a}
    \left(
      c_{j,b}^{\dagger}c_{j+1,b}
      +c_{j+1,b}^{\dagger}c_{j,b}
    \right),
  \end{split}
\end{equation}
which verifies Eq.~\eqref{eq:hopping_hamiltonian} link by link. Because the transformation leaves every $\hat n_{j,b}$ invariant, it also leaves $H_m$, $D_n$, $D_n^2$, and all flavor charges unchanged.

\subsection{Flavor-Chain Jordan-Wigner Transformation}

For $\alpha\in\{I,R\}$, define
\begin{equation}
  f_{j,b}^{(I)}\equiv\chi_{j,b},
  \qquad
  f_{j,b}^{(R)}\equiv c_{j,b}.
\end{equation}
We apply the same flavor-chain Jordan-Wigner ordering separately in each convention:
\begin{equation}
  \begin{aligned}
    f_{j,b}^{(\alpha)}
    &=\mathcal K_b^{(\alpha)}
    \left(
      \prod_{\ell=0}^{j-1}Z_{q(\ell,b)}^{(\alpha)}
    \right)
    \\
    &\quad\times
    \frac{
      X_{q(j,b)}^{(\alpha)}
      +iY_{q(j,b)}^{(\alpha)}
    }{2},
    \\
    \mathcal K_b^{(\alpha)}
    &=\prod_{b'=0}^{b-1}
    \prod_{\ell=0}^{2\LD-1}
    Z_{q(\ell,b')}^{(\alpha)}.
  \end{aligned}
  \label{eq:generic_jw_string}
\end{equation}
The parity string in the first line follows increasing $j$ within a fixed flavor chain. The Klein factor $\mathcal K_b^{(\alpha)}$ orders the flavor chains algebraically and enforces anticommutation between distinct flavors. It cancels from Hamiltonian terms containing an even number of operators from every participating flavor. This algebraic ordering is independent of the site-major register index assigned to each qubit.

The relation $f_{j,b}^{(I)}=i^j f_{j,b}^{(R)}$ is implemented by an on-site $Z$ rotation at each staggered site. It rotates the local $X$ and $Y$ axes while leaving the occupation operator invariant:
\begin{equation}
  Z_{q(j,b)}^{(I)}
  =Z_{q(j,b)}^{(R)}
  \equiv Z_{q(j,b)}.
\end{equation}
The mass, interaction, and flavor charges therefore have the same Pauli-$Z$ representation in both conventions.

For neighboring sites on the same flavor chain, the intermediate parity string is empty. Equation~\eqref{eq:generic_jw_string} gives
\begin{equation}
  \begin{aligned}
    &-i\left(
      f_{j,b}^{(I)\dagger}f_{j+1,b}^{(I)}
      -f_{j+1,b}^{(I)\dagger}f_{j,b}^{(I)}
    \right)
    \\
    &\qquad=
    \frac{1}{2}
    \left[
      X_{q(j,b)}^{(I)}Y_{q(j+1,b)}^{(I)}
      -Y_{q(j,b)}^{(I)}X_{q(j+1,b)}^{(I)}
    \right],
    \\
    &f_{j,b}^{(R)\dagger}f_{j+1,b}^{(R)}
    +f_{j+1,b}^{(R)\dagger}f_{j,b}^{(R)}
    \\
    &\qquad=
    \frac{1}{2}
    \left[
      X_{q(j,b)}^{(R)}X_{q(j+1,b)}^{(R)}
      +Y_{q(j,b)}^{(R)}Y_{q(j+1,b)}^{(R)}
    \right].
  \end{aligned}
\end{equation}
Together with $\hat n_{j,b}=(1-Z_{q(j,b)})/2$, these identities give the full Pauli hopping operators summarized in Eq.~\eqref{eq:hopping_pauli_comparison} and the Pauli form of the staggered mass term.

\subsection{Interaction Decomposition and Identity Offset}

Define the flavor-resolved scalar-density difference
\begin{equation}
  \Delta_{n,b}
  =\hat n_{2n,b}-\hat n_{2n+1,b}.
\end{equation}
Then
\begin{equation}
  \begin{split}
    D_n^2
    ={}&\sum_{b=0}^{\Nf-1}\Delta_{n,b}^2
    ~+~~2\sum_{0\leq b<b'\leq\Nf-1}
    \Delta_{n,b}\Delta_{n,b'}.
  \end{split}
  \label{eq:dn_expansion}
\end{equation}
The first term is flavor diagonal, whereas the second contains the cross-flavor density-density couplings. In the qubit representation,
\begin{equation}
  \begin{aligned}
    \Delta_{n,b}
    &=\frac{
      Z_{q(2n+1,b)}-Z_{q(2n,b)}
    }{2},
    \\
    \Delta_{n,b}^2
    &=\frac{1}{2}
    \left[
      I-Z_{q(2n,b)}Z_{q(2n+1,b)}
    \right].
  \end{aligned}
\end{equation}
Substitution into Eq.~\eqref{eq:interaction_hamiltonian} gives
\begin{equation}
  \begin{aligned}
    H_{\mathrm{int}}
    ={}&-\frac{g^2\LD\Nf}{4a}
    +\frac{g^2}{4a}
    \sum_{n=0}^{\LD-1}
    \sum_{b=0}^{\Nf-1}
    Z_{q(2n,b)}Z_{q(2n+1,b)}
    \\
    &-\frac{g^2}{a}
    \sum_{n=0}^{\LD-1}
    \sum_{0\leq b<b'\leq\Nf-1}
    \Delta_{n,b}\Delta_{n,b'}.
  \end{aligned}
\end{equation}
The first term is the interaction identity offset displayed in the main text, while the remaining terms contain the same-flavor and cross-flavor Pauli interactions.

\section{Conservation Algebra and Fixed-Circuit Schedules}

\label{app:fixed_ansatze}

\subsection{Pair-number conservation}

For a qubit pair \((u,v)\), define
\begin{equation}
  \begin{aligned}
    N_{uv}
    &=
    \hat n_u+\hat n_v
    =
    \frac{2I-Z_u-Z_v}{2},
    \\
    A_{uv}&=X_uX_v+Y_uY_v .
  \end{aligned}
  \label{eq:fixed_pair_operators}
\end{equation}
Because \(X_uX_v\) and \(Y_uY_v\) commute,
\begin{equation}
  R_{Y_uY_v}(\alpha)R_{X_uX_v}(\alpha)
  =
  e^{-i\alpha A_{uv}}.
  \label{eq:fixed_exchange_product}
\end{equation}
The tied product, rather than either factor separately, preserves pair occupation:
\begin{equation}
  \begin{aligned}
    \bigl[A_{uv},N_{uv}\bigr]
    &=
    [Z_uZ_v,N_{uv}]
    \\
    &=
    [Z_u,N_{uv}]
    =
    [Z_v,N_{uv}]
    =0 .
  \end{aligned}
  \label{eq:fixed_number_commutators}
\end{equation}
Thus the chronological action of \(B\) is an \(XX\) rotation, the equal-angle \(YY\) rotation, a \(ZZ\) rotation, and independent rotations about \(Z_u\) and \(Z_v\). Every complete block preserves \(N_{uv}\). For endpoints belonging to flavors \(b\ne c\), the exchange component commutes with \(N_b+N_c\) but not generally with \(N_b\) or \(N_c\) separately.

For the flavor-number-conserving ansatz, the \(B\) blocks, which have the \(XX/YY\) product, are used only when both endpoints belong to the same flavor, ensuring there is no redistribution of occupation between different flavors. The chronological action of \(D\) is a \(ZZ\) rotation followed by the \(+Z_u\) and \(-Z_v\) phases. Since every factor is diagonal, this block also commutes with every flavor charge when \(u\) and \(v\) belong to different flavors.

\subsection{Edge classes and ordered layers}

For each unordered flavor pair $(b<c)$, the four translated cross-flavor edge classes are
\begin{equation}
  \begin{aligned}
    e_{bc}^{st}(n)
    &=
    \bigl(q(2n+s,b),q(2n+t,c)\bigr),
    \\
    (s,t)&\in\{(0,0),(1,1),(0,1),(1,0)\},
    \\
    0&\le n\le\LD-1,
    \qquad 0\le b<c\le\Nf-1.
  \end{aligned}
  \label{eq:fixed_cross_flavor_topologies}
\end{equation}

For each flavor \(b\), define the ordered edge lists
\begin{align}
  \mathcal E_{\mathrm{nn}}^{b}
  &=
  \bigl(
    (q(j,b),q(j+1,b))
  \bigr)_{j=0}^{2\LD-2},
  \label{eq:fixed_nn_edges}\\
  \mathcal E_{\mathrm{even}}^{b}
  &=
  \bigl(
    (q(2n,b),q(2n+1,b))
  \bigr)_{n=0}^{\LD-1},
  \notag\\
  \mathcal E_{\mathrm{odd}}^{b}
  &=
  \bigl(
    (q(2n+1,b),q(2n+2,b))
  \bigr)_{n=0}^{\LD-2},
  \label{eq:fixed_parity_edges}\\
  \mathcal E_{\mathrm{cell}}^{b}
  &=
  \bigl(
    (q(2n,b),q(2n+1,b))
  \bigr)_{n=0}^{\LD-1}.
  \label{eq:fixed_cell_edges}
\end{align}
The entries of each list are written in gate-application order. For an ordered list \(\mathcal E=(e_1,\ldots,e_M)\), a block family \(B\), and a shared tuple \(\lambda\), define
\begin{equation}
  \mathcal P[B;\mathcal E,\lambda]
  =
  B_{e_M}(\lambda)\cdots B_{e_2}(\lambda)B_{e_1}(\lambda),
  \label{eq:fixed_ordered_product}
\end{equation}
so that the edge \(e_1\) is acted on first.

The cross-flavor edges are the \(e_{bc}^{st}(n)\) defined in Eq.~\eqref{eq:fixed_cross_flavor_topologies}. For each unordered flavor pair \(b<c\) and topology \((s,t)\), the tuple \(\lambda_{bc,st}\) is shared over \(n\), while distinct flavor pairs and topologies retain independent tuples.

Introduce the per-flavor products
\begin{equation}
  \begin{aligned}
    Q_{p,b}
    &=
    \mathcal P[
      B;\mathcal E_p^{b},
      \lambda_{\mathrm{nn},b}
    ],
    \quad p\in\{\mathrm{even},\mathrm{odd}\}.
  \end{aligned}
  \label{eq:fixed_per_flavor_products}
\end{equation}
For compactness, define \(\overleftarrow{\prod}_{b=0}^{\Nf-1}Q_b\equiv Q_{\Nf-1}\cdots Q_0\), so that flavor \(b=0\) acts first. The nearest-neighbor schedule is thus
\begin{align}
  U_{\mathrm{nn}}
  ={}&
  \left(
    \overleftarrow{\prod}_{b=0}^{\Nf-1}Q_{\mathrm{odd},b}
  \right)
  \notag 
  \times
  \left(
    \overleftarrow{\prod}_{b=0}^{\Nf-1}Q_{\mathrm{even},b}
  \right).
  \label{eq:fixed_brickwall_schedule}
\end{align}
Thus the total-number stage and the flavor-number stage both traverse all even-start lists in increasing \((b,n)\), followed by all odd-start lists in the same order, using the same tuple in both parity sweeps. Chronologically, the additional intracell stage follows the nearest-neighbor stage and traverses increasing \((b,n)\) with a distinct tuple \(\lambda_{\mathrm{cell},b}\). The cross-flavor stage is last: it traverses increasing \(b\), then \(c>b\), then \(n\), with within-cell order \((0,0),(1,1),(0,1),(1,0)\). Parameters are shared only within a fixed translated topology class.

\subsection{Independent and anti-tied local phases}

If the phases in $B$ were independent, they would decompose as
\begin{equation}
  \begin{aligned}
    \gamma Z_u+\delta Z_v
    &=
    \frac{\gamma-\delta}{2}D_{uv}
    +
    \frac{\gamma+\delta}{2}S_{uv},
    \\
    D_{uv}&=Z_u-Z_v,
    \\ 
    S_{uv}&=Z_u+Z_v .
  \end{aligned}
  \label{eq:fixed_phase_decomposition}
\end{equation}
The anti-tied choice \(\delta=-\gamma\) retains only \(D_{uv}\). With
\begin{equation}
  C_{uv}=X_uY_v-Y_uX_v,
\end{equation}
the local exchange algebra is
\begin{equation}
  [D_{uv},A_{uv}]=-4iC_{uv},
  \qquad
  [S_{uv},A_{uv}]=0 .
  \label{eq:fixed_phase_commutators}
\end{equation}
Thus, the relative phase is conjugate to the exchange direction and can rotate \(A_{uv}\) into a current-like direction. By contrast, \(S_{uv}=2I-2N_{uv}\) would be diagonal in pair occupation and vanish within the one-particle pair sector. 

\section{SC-ADAPT Operator Algebra and Circuit Schedules}

\label{app:sc_adapt_details}

\subsection{Parent Operators and Geometric Pool}

The volume parents and their Hamiltonian normalizations are
\begin{equation}
  \begin{aligned}
    \Theta_m^V
    &=\sum_{j=0}^{2\LD-1}\sum_{b=0}^{\Nf-1}
    (-1)^j\hat n_{j,b},
    \\
    \Theta_h^V(d)
    &=\frac12\sum_{j=0}^{2\LD-1-d}\sum_{b=0}^{\Nf-1}
    \left(c_{j,b}^{\dagger}c_{j+d,b}+\hc\right),
    \\
    \Theta_g^V
    &=\sum_{n=0}^{\LD-1}D_n^2,
  \end{aligned}
  \label{eq:scalable_building_blocks}
\end{equation}
\begin{equation}
  H_m=m\Theta_m^V,
  ~~
H_{\mathrm{hop}}^{(R)}=\frac1a\Theta_h^V(1),
~~
  H_{\mathrm{int}}=-\frac{g^2}{2a}\Theta_g^V.
  \label{eq:sc_parent_hamiltonians}
\end{equation}
The staggered identity contribution to \(H_m\) cancels over the even number of sites. Each parent commutes with every \(N_b\), so their commutators preserve all flavor numbers.

The two surface families used in the main text are
\begin{equation}
  \begin{aligned}
    O_{mh}^{S0}(d)
    &=\frac12\sum_{b=0}^{\Nf-1}
    \left[J_{0,b}^{(d)}+J_{2\LD-1-d,b}^{(d)}\right],
    \\
    O_{mh}^{S1}(d)
    &=-\frac12\sum_{b=0}^{\Nf-1}
    \left[J_{1,b}^{(d)}+J_{2\LD-2-d,b}^{(d)}\right].
  \end{aligned}
  \label{eq:surface_omh}
\end{equation}
They retain the boundary structures and staggered relative sign of the volume current, with half its local coefficient magnitude \cite{PRXQuantum.5.020315}.

The available odd separations are
\begin{equation}
  \begin{aligned}
    \mathcal D_0
    &=\{d\in2\mathbb Z+1:1\le d\le2\LD-3\},
    \\
    \mathcal D_1
    &=\{d\in2\mathbb Z+1:1\le d\le2\LD-5\},
  \end{aligned}
  \label{eq:surface_ranges}
\end{equation}
where \(\mathcal D_0\) applies to \(O_{mh}^{V}\), \(O_{mh}^{S0}\), and \(O_{gh}^{V}\), and \(\mathcal D_1\) to \(O_{mh}^{S1}\). Before the numerical caps in Eq.~\eqref{eq:numerical_pool}, the geometric pool is
\begin{equation}
  \begin{aligned}
    \mathcal P_{\mathrm{geom}}
    ={}&\left\{O_{mh}^{V}(d),O_{mh}^{S0}(d),O_{gh}^{V}(d)\right\}_{d\in\mathcal D_0}
    \\
    &{}\cup\left\{O_{mh}^{S1}(d)\right\}_{d\in\mathcal D_1}.
  \end{aligned}
  \label{eq:geometric_pool}
\end{equation}

\subsection{Current Strings and Mass-Hopping Reductions}

\label{sec:mass_hopping}

In the convention-\(R\) Pauli frame, define the intermediate flavor-chain parity string
\begin{equation}
  \mathcal Z_{j,b}^{(d)}
  =
  \prod_{\ell=j+1}^{j+d-1}Z_{q(\ell,b)},
  \qquad
  \mathcal Z_{j,b}^{(1)}=I.
  \label{eq:sc_jw_string}
\end{equation}
The two current strings and their antisymmetric combination are
\begin{equation}
  \begin{aligned}
    P_{XY;j,b}^{(d)}
    &=
    X_{q(j,b)}\mathcal Z_{j,b}^{(d)}
    \\
    &\quad\times Y_{q(j+d,b)},
    \\
    P_{YX;j,b}^{(d)}
    &=
    Y_{q(j,b)}\mathcal Z_{j,b}^{(d)}
    \\
    &\quad\times X_{q(j+d,b)},
    \\
    K_{j,b}^{(d)}
    &=P_{XY;j,b}^{(d)}-P_{YX;j,b}^{(d)}.
  \end{aligned}
  \label{eq:current_string}
\end{equation}
The Jordan-Wigner transformation in Eq.~\eqref{eq:generic_jw_string} gives
\begin{equation}
  J_{j,b}^{(d)}=-\frac12K_{j,b}^{(d)}.
  \label{eq:current_jw_map}
\end{equation}
For arbitrary \(d\ge1\), direct evaluation of the mass-hopping commutator yields
\begin{equation}
  \begin{aligned}
    i[\Theta_m^V,\Theta_h^V(d)]
    ={}&
    \frac{i}{2}
    \sum_{j=0}^{2\LD-1-d}
    \sum_{b=0}^{\Nf-1}
    \\
    &\times
    \left[(-1)^j-(-1)^{j+d}\right]
    \\
    &\times
    \left(
      c_{j,b}^{\dagger}c_{j+d,b}
      -c_{j+d,b}^{\dagger}c_{j,b}
    \right).
  \end{aligned}
\end{equation}
This expression vanishes for even \(d\). For odd \(d\), Eqs.~\eqref{eq:omh_definition} and~\eqref{eq:current_jw_map} give
\begin{equation}
  O_{mh}^V(d)
  =
  -\frac12
  \sum_{j=0}^{2\LD-1-d}
  \sum_{b=0}^{\Nf-1}
  (-1)^jK_{j,b}^{(d)}.
  \label{eq:omh_pauli}
\end{equation}
Similarly, Eq.~\eqref{eq:surface_omh} becomes
\begin{equation}
  \begin{aligned}
    O_{mh}^{S0}(d)
    &=
    -\frac14\sum_{b=0}^{\Nf-1}
    \left[
      K_{0,b}^{(d)}+K_{2\LD-1-d,b}^{(d)}
    \right],
    \\
    O_{mh}^{S1}(d)
    &=
    +\frac14\sum_{b=0}^{\Nf-1}
    \left[
      K_{1,b}^{(d)}+K_{2\LD-2-d,b}^{(d)}
    \right].
  \end{aligned}
\end{equation}

\subsection{Cross-Flavor Interaction-Hopping Reduction}

\label{sec:interaction_hopping}

Let
\begin{equation}
  \begin{aligned}
    \sigma_n(j,d)
    ={}&
    \delta_{2n,j}-\delta_{2n+1,j}
    \\
    &-
    \delta_{2n,j+d}+\delta_{2n+1,j+d}.
  \end{aligned}
  \label{eq:sigma_overlap}
\end{equation}
For a current of flavor \(\beta\), the cross-flavor part of the commutator with \(D_n^2\) retains the density differences \(\Delta_{n,b}\) with \(b\ne\beta\). The projected generator in Eq.~\eqref{eq:ogh_definition} is therefore
\begin{equation}
  \begin{aligned}
    O_{gh}^V(d)
    ={}&
    \sum_{n=0}^{\LD-1}
    \sum_{j=0}^{2\LD-1-d}
    \sum_{\beta=0}^{\Nf-1}
    \sigma_n(j,d)
    \\
    &\times
    \sum_{\substack{b=0\\b\ne\beta}}^{\Nf-1}
    \Delta_{n,b}J_{j,\beta}^{(d)}.
  \end{aligned}
  \label{eq:ogh_fermion}
\end{equation}
Using
\(\Delta_{n,b}=[Z_{q(2n+1,b)}-Z_{q(2n,b)}]/2\)
and Eq.~\eqref{eq:current_jw_map} gives
\begin{equation}
  \begin{aligned}
    O_{gh}^V(d)
    ={}&
    -\frac14
    \sum_{n=0}^{\LD-1}
    \sum_{j=0}^{2\LD-1-d}
    \sum_{\beta=0}^{\Nf-1}
    \sigma_n(j,d)
    \\
    &\times
    \sum_{\substack{b=0\\b\ne\beta}}^{\Nf-1}
    \left[
      Z_{q(2n+1,b)}-Z_{q(2n,b)}
    \right]
    K_{j,\beta}^{(d)}.
  \end{aligned}
  \label{eq:ogh_pauli}
\end{equation}
Each factor \(\Delta_{n,b}J_{j,\beta}^{(d)}\) preserves every flavor number. The projection is defined in the fixed flavor basis and enforces the Abelian charges $\{N_b\}$, The symmetry constraint imposed by this projection is conservation of the Abelian charges $\{N_b\}$ in the chosen flavor basis.

\subsection{Exact Current Blocks}

\label{sec:trotter_circuit}

For an ordered qubit pair \((u,v)\), we adopt the circuit convention of Fig.~3 of Ref.~\cite{PRXQuantum.5.020315} and define
\begin{equation}
  \begin{aligned}
    R_{\pm}^{uv}(\varphi)
    &=
    \exp\!\left[
      -\frac{i\varphi}{2}
      \left(Y_uX_v\pm X_uY_v\right)
    \right],
    \\
    \kappa_{uv}&=X_uY_v-Y_uX_v.
  \end{aligned}
  \label{eq:rpm_definition}
\end{equation}
In particular,
\begin{equation}
  R_-^{uv}(\varphi)
  =
  \exp\!\left(\frac{i\varphi}{2}\kappa_{uv}\right).
  \label{eq:rminus_current}
\end{equation}
With \(R_y(\varphi)=e^{-i\varphi Y/2}\) and \(R_z(\varphi)=e^{-i\varphi Z/2}\), the gate-application sequence is listed from left to right:
\begin{equation}
  \begin{gathered}
    S_u,H_u;\qquad Z_v,H_v,S_v;
    \\
    \operatorname{CNOT}_{u\rightarrow v};
    \\
    R_y^{(u)}(\pm\varphi),\ R_z^{(v)}(\varphi);
    \\
    \operatorname{CNOT}_{u\rightarrow v};
    \\
    H_u,S_u^{\dagger};\qquad
    S_v^{\dagger},H_v,Z_v.
  \end{gathered}
  \label{eq:rpm_gate_sequence}
\end{equation}
The positive sign in the \(R_y\) angle gives \(R_+\), and the negative sign gives \(R_-\).

Extending the odd-range X patterns displayed in Ref.~\cite{PRXQuantum.5.020315} in this convention, for odd \(d=2r+1\) we use the following ordered construction. Let \(q_k=q(j+k,b)\) for \(k=0,\ldots,d\). The forward legs apply, for \(\ell=0,\ldots,r-1\) in increasing order,
\begin{equation}
  \begin{aligned}
    &R_+^{q_\ell q_{\ell+1}}(-\pi/2),
    \\
    &R_+^{q_{d-\ell-1}q_{d-\ell}}(-\pi/2),
  \end{aligned}
  \label{eq:xpattern_forward}
\end{equation}
followed by \(R_-^{q_rq_{r+1}}(s_d\alpha)\), where
\begin{equation}
  s_d=(-1)^r=(-1)^{(d-1)/2}.
  \label{eq:xpattern_sign}
\end{equation}
The forward legs are then undone in reverse radial order using \(R_+(+\pi/2)\). If \(A_{j,b}^{(d)}\) denotes their ordered product, the complete X-pattern block satisfies the exact identity
\begin{equation}
  \begin{aligned}
    W_{j,b}^{(d)}(\alpha)
    &=
    A_{j,b}^{(d)\dagger} \, 
    R_-^{q_rq_{r+1}}(s_d\alpha) \, 
    A_{j,b}^{(d)}
    \\
    &=
    \exp\!\left[
      \frac{i\alpha}{2}K_{j,b}^{(d)}
    \right].
  \end{aligned}
  \label{eq:xpattern_block}
\end{equation}
Conjugating the central current through the ordered legs yields \(s_d=(-1)^{(d-1)/2}\), consistent with the odd-range patterns displayed in Ref.~\cite{PRXQuantum.5.020315}. This factor corrects the X-pattern orientation and is distinct from the staggered coefficient \((-1)^j\) in \(O_{mh}^V(d)\).

\subsection{Mass-Hopping Product Schedules}

Let
\(\mathcal L_V(d)=((j_1,b_1),\ldots,(j_{T_V^{(d)}},b_{T_V^{(d)}}))\)
denote the volume pairs in gate-application order. Define
\begin{equation}
  B_{V,t}^{(d)}(\vartheta)
  =
  W_{j_t,b_t}^{(d)}
  \!\left((-1)^{j_t}\vartheta\right).
\end{equation}
For the surface families, define the right-boundary start sites
\begin{equation}
  \ell_0=2\LD-1-d,
  \qquad
  \ell_1=2\LD-2-d,
\end{equation}
and let \(t=1,\ldots,2\Nf\), with
\begin{equation}
  B_{S0,t}^{(d)}(\vartheta)
  =
  \begin{cases}
    W_{0,t-1}^{(d)}(\vartheta/2),
      &1\le t\le\Nf,\\
    W_{\ell_0,t-\Nf-1}^{(d)}(\vartheta/2),
      &\Nf<t\le2\Nf,
  \end{cases}
\end{equation}
and
\begin{equation}
  B_{S1,t}^{(d)}(\vartheta)
  =
  \begin{cases}
    W_{1,t-1}^{(d)}(-\vartheta/2),
      &1\le t\le\Nf,\\
    W_{\ell_1,t-\Nf-1}^{(d)}(-\vartheta/2),
      &\Nf<t\le2\Nf.
  \end{cases}
\end{equation}
Thus the left boundary is applied before the right boundary, with increasing flavor index at each boundary. Under standard rightmost-first operator action, the scheduled candidate is
\begin{equation}
  \begin{aligned}
    \widetilde U_X^{(d)}(\vartheta)
    ={}&
    B_{X,T_X^{(d)}}^{(d)}(\vartheta) \times
    \cdots
    \times
    B_{X,2}^{(d)}(\vartheta)
    B_{X,1}^{(d)}(\vartheta),
    \\
    &X\in\{V,S0,S1\}.
  \end{aligned}
  \label{eq:mass_hopping_schedules}
\end{equation}
With the formal signs in Eqs.~\eqref{eq:omh_pauli} and~\eqref{eq:surface_omh},
\begin{equation}
  \begin{aligned}
    \widetilde U_X^{(d)}(\vartheta)
    &=
    e^{-i\vartheta O_{mh}^{X}(d)}
    +\mathcal O(\vartheta^2),
    \\
    X&\in\{V,S0,S1\}.
  \end{aligned}
  \label{eq:mass_hopping_parameter_map}
\end{equation}
The common minus sign is absorbed by \(\theta=-\vartheta\). The factor \(1/2\) in the surface angles implements the relative normalization in Eq.~\eqref{eq:surface_omh} and is retained in append-point gradient comparisons.

For the volume family, blocks are grouped by increasing flavor index. Within each flavor, all even starts precede all odd starts for \(d=1\) and \(d=3\). For \(d>3\), the residue layers have stride \(d-1=2r\) and are ordered
\begin{equation}
  0,r,1,r+1,\ldots,r-1,2r-1,
  \label{eq:volume_layer_order}
\end{equation}
with increasing start index within each layer. These fixed orders are part of the finite-depth candidate circuit.

\subsection{Density-Dressed Blocks and Adaptive-Circuit Order}

Let \(q_c=q(\ell,b)\) be a cross-flavor density qubit and let the current of flavor \(\beta\ne b\) begin at \(q_0=q(j,\beta)\). Controlled-\(Z\) conjugation gives
\begin{equation}
  \begin{aligned}
    &\operatorname{CZ}_{q_c,q_0}
    K_{j,\beta}^{(d)} \, 
    \operatorname{CZ}_{q_c,q_0}
=    Z_{q_c}K_{j,\beta}^{(d)},
  \end{aligned}
  \label{eq:cz_dressed_current}
\end{equation}
and hence the exact dressed block
\begin{equation}
  \begin{aligned}
    &\operatorname{CZ}_{q_c,q_0}
    W_{j,\beta}^{(d)}(\alpha)
    \operatorname{CZ}_{q_c,q_0}
=
    \exp\!\left[
      \frac{i\alpha}{2}
      Z_{q_c}K_{j,\beta}^{(d)}
    \right].
  \end{aligned}
  \label{eq:cz_dressed_rotation}
\end{equation}
Write Eq.~\eqref{eq:ogh_pauli} as the signed term list in gate-application order,
\begin{equation}
  O_{gh}^V(d)
  =
  -\frac14
  \sum_{t=1}^{T_{gh}^{(d)}}s_tZ_tK_t,
  \qquad
  s_t\in\{-1,+1\},
  \label{eq:ogh_term_list}
\end{equation}
where repeated entries retain algebraic multiplicities. For each listed term, define
\begin{equation}
  \begin{aligned}
    D_t^{(d)}(\vartheta)
    &={}
    \operatorname{CZ}_{q_{c,t},q_{0,t}}
    W_t^{(d)}
    \!\left(\eta_{gh}s_t\vartheta\right)
    \, \operatorname{CZ}_{q_{c,t},q_{0,t}},
    \\
    \eta_{gh}&=0.2.
  \end{aligned}
  \label{eq:ogh_weight}
\end{equation}
The implemented interaction candidate is
\begin{equation}
  \begin{aligned}
    \widetilde U_{Vgh}^{(d)}(\vartheta)
    ={}&
    D_{T_{gh}^{(d)}}^{(d)}(\vartheta)
    \cdots
    D_2^{(d)}(\vartheta)D_1^{(d)}(\vartheta)
    \\
    ={}&
    e^{-2i\eta_{gh}\vartheta O_{gh}^V(d)}
    +\mathcal O(\vartheta^2).
  \end{aligned}
  \label{eq:ogh_parameter_map}
\end{equation}
Thus the implemented append-point generator is $-2\eta_{gh}O_{gh}^V(d)=-0.4O_{gh}^V(d)$ for the value $\eta_{gh}=0.2$ used in the reported calculations. Its overall sign is absorbable into an unrestricted optimized parameter, while its magnitude rescales the raw append-point score relative to the other families and can therefore affect the selected sequence. The coefficient $\eta_{gh}$ is a dimensionless pool-normalization, or gradient-preconditioning, hyperparameter introduced to reduce the raw-score scale of the larger density-dressed family; it is distinct from the Hamiltonian coupling $g^2$. We use $\eta_{gh}=0.2$ for the benchmark circuits.

The interaction factors are ordered by physical cell, hopping flavor, distinct dressing flavor, active density site (odd then even), dressing-density endpoint (odd then even), and current endpoint (left then right when present). If \((\widetilde U_1,\ldots,\widetilde U_M)\) is the selected sequence in append order, then
\begin{equation}
  \begin{aligned}
    |\psi_M\rangle
    ={}&
    \widetilde U_M\cdots\widetilde U_2
   \,  \widetilde U_1|\Phi_0\rangle.
  \end{aligned}
  \label{eq:selected_circuit_order}
\end{equation}

\begin{table*}[t]
  \centering
  \footnotesize
  \caption{Representative raw-score SC-ADAPT selection sequence for $\Nf=2$, $\LD=4$, $m=0.5$, $g^2=0.4$, $\eta_{gh}=0.2$, and $\delta=10^{-3}$. The selected and runner-up columns give the candidates with the two largest append-point magnitudes before optimization. The gap is their absolute score difference, $E_k$ is the reoptimized energy, and $r_k$ is the maximum post-optimization raw score within the retained pool.}
  \label{tab:sc_adapt_selection_audit}
  \setlength{\tabcolsep}{4pt}
  \renewcommand{\arraystretch}{1.15}
  \begin{tabular*}{\linewidth}{@{\extracolsep{\fill}}ccrcrrr@{}}
    \hline\hline
    \multicolumn{1}{c}{$k$} &
    \multicolumn{1}{c}{Selected} &
    \multicolumn{1}{c}{$|g_{\mathrm{sel}}^{\mathrm{FD}}|$} &
    \multicolumn{1}{c}{Runner-up} &
    \multicolumn{1}{c}{Gap} &
    \multicolumn{1}{c}{$E_k$} &
    \multicolumn{1}{c}{$r_k$} \\
    \hline
    1 & $V(1)$       & $13.99997$ & $V_{gh}(1)$ & $2.80001$               & $-8.78932961$ & $0.875541$ \\
    2 & $V(3)$       & $0.875541$ & $V_{gh}(1)$ & $0.147044$              & $-8.79745391$ & $0.784304$ \\
    3 & $V_{gh}(1)$  & $0.784304$ & $S_1(1)$    & $0.604139$              & $-8.80864163$ & $0.175621$ \\
    4 & $S_1(1)$     & $0.175621$ & $S_0(1)$    & $7.60634\times10^{-3}$ & $-8.81366680$ & $0.116657$ \\
    5 & $S_0(1)$     & $0.116657$ & $S_1(3)$    & $1.91951\times10^{-2}$ & $-8.81676878$ & $0.0920309$ \\
    6 & $S_1(3)$     & $0.0920309$ & $S_0(3)$   & $2.93527\times10^{-2}$ & $-8.81893050$ & $0.0363964$ \\
    7 & $V(5)$       & $0.0363964$ & $V_{gh}(3)$ & $1.23032\times10^{-3}$ & $-8.81895417$ & $0.0355096$ \\
    \hline\hline
  \end{tabular*}
\end{table*}

Each \(R_\pm\) primitive, complete X-pattern block \(W\), and controlled-\(Z\)-dressed \(W\) block is exact. First-order product error enters only when the exponential of a sum of translated or overlapping blocks is replaced by the fixed ordered product. The complete current and dressed-current blocks preserve every \(N_b\), although their internal basis-change gates need not do so individually, and the resulting candidate circuits remain real in the occupation basis.

\subsection{Representative SC-ADAPT Selection Scores}

For the $\Nf=2$, $\LD=4$ calculation discussed in Sec.~\ref{sec:sc_adapt_selection_audit}, define
\begin{equation}
  r_k=\max_{i\in\mathcal P}|g_{i,k}^{\mathrm{FD}}|,
  \label{eq:raw_pool_maximum}
\end{equation}
where the score is evaluated after reoptimization at step $k$. This is a maximum over the retained pool $\mathcal P$, not a Hilbert-space residual, and its magnitude depends on the candidate coordinates. Table~\ref{tab:sc_adapt_selection_audit} reports the two leading pre-append candidates, the selected score, and their score gap at each step.

\subsection{Finite-Volume Sequence Coordinates}

For the one-flavor transfer test, the frozen append order is the $\Nf=1$, $\LD=12$-15 sequence in Table~\ref{tab:sc_adapt_operator_stabilization}, with the tabulated $\LD=12$ coordinate vector
\begin{equation}
  \begin{aligned}
    \boldsymbol\theta_{12}^{\mathrm{tab}}
    ={}&\bigl(-0.3065,0.0384,-0.0058,\\
       &-0.1179,0.0015,0.0530,-0.0527,\\
       &0.0514,0.0113,-0.0004\bigr).
  \end{aligned}
  \label{eq:nf1_transfer_coordinates}
\end{equation}
The two occurrences of $V(1)$ are distinct append coordinates.

The optimized parameters discussed here are coordinates of the implemented fixed factorization rather than uniformly normalized coefficients of the formal generators; the \(V_{gh}\) coordinates include the fixed \(\eta_{gh}=0.2\) scaling. Within the one-flavor stability window \(\LD=12\)-15, the leading \(V(1)\) coordinate changes from \(-0.3065\) to \(-0.3043\), while the second \(V(1)\) occurrence changes from \(0.0113\) to \(0.0095\). For two flavors, from \(\LD=4\) to 7, the leading \(V(1)\) coordinate changes from \(-0.25019\) to \(-0.24506\), and the \(V_{gh}(1)\) coordinate from \(-0.02857\) to \(-0.02649\). These changes are small but nonzero and do not establish an infinite-volume limit.

The one-flavor $V(9)$ generator first appears at $\LD=8$, while $V(7)$ occupies the eighth position at $\LD=7$, the sixth at $\LD=8$-11, and the fifth at $\LD=12$-15. The origin of these finite-volume changes in operator order remains unresolved. Finite-volume interchange among retained operators was also observed in the Schwinger-model study \cite{PRXQuantum.5.020315}.

\bibliography{reference}

\end{document}